\documentclass[Journal,letterpaper,NoLineNumbers]{ascelike-new}
\usepackage[utf8]{inputenc}
\usepackage[T1]{fontenc}
\usepackage{lmodern}
\usepackage{graphicx}
\usepackage[figurename=Fig.,labelfont=bf,labelsep=period]{caption}
\usepackage{amsmath}
\usepackage{newtxtext,newtxmath}
\usepackage[colorlinks=true,citecolor=red,linkcolor=black]{hyperref}
\usepackage{nameref}
\usepackage{bm}
\usepackage{mathrsfs}
\usepackage{amsfonts}
\usepackage{algorithm}
\usepackage[noend]{algpseudocode}
\usepackage{color}
\usepackage{setspace}
\usepackage{import}
\usepackage{url}
\usepackage{multicol}
\usepackage{subfigure}
\usepackage{amsmath,bbm,mathtools,bm}
\usepackage{multicol}
\usepackage{multirow}
\usepackage{xcolor}

\definecolor{officegreen}{rgb}{0.0, 0.5, 0.0}

\def\omg{{\Omega}}

\newcommand{\mcD}{\mathcal{D}}

\newcommand{\mcN}{\mathcal{N}}
\newcommand{\mcL}{\mathcal{L}}

\def \Cb{\mathbf{C}}

\def \vb{\mathbf{v}}

\def \real{\mathbb{R}}
\def \R{\mathbb{R}}

\newcommand{\verti}[1]{{\left\vert #1
    \right\vert}}  
    
\newcommand{\vertii}[1]{{\left\vert\left\vert #1
    \right\vert\right\vert}}

\usepackage[symbol]{footmisc}
\renewcommand{\thefootnote}{\fnsymbol{footnote}}

\NameTag{Fan, \today}

\begin{document}

\title{Bayesian Nonlocal Operator Regression (BNOR): A Data-Driven Learning Framework of Nonlocal Models with Uncertainty Quantification}

\author[1]{Yiming Fan}
\author[2]{Marta D'Elia}
\author[1,*]{Yue Yu}
\author[2]{Habib N. Najm}
\author[3]{Stewart Silling}

\affil[1]{Department of Mathematics, Lehigh University, Bethlehem, PA 18015, USA}
\affil[2]{Sandia National Laboratories, Livermore, CA}
\affil[3]{Center for Computing Research, Sandia National Laboratories, Albuquerque, NM}
\affil[*]{Corresponding author: Yue Yu, Email: yuy214@lehigh.edu}

\maketitle

\begin{abstract}
We consider the problem of modeling heterogeneous materials where micro-scale dynamics and interactions affect global behavior. In the presence of heterogeneities in material microstructure it is often impractical, if not impossible, to provide quantitative characterization of material response. The goal of this work is to develop a Bayesian framework for uncertainty quantification (UQ) in material response prediction when using nonlocal models. Our approach combines the nonlocal operator regression (NOR) technique and Bayesian inference. Specifically, we use a Markov chain Monte Carlo (MCMC) method to sample the posterior probability distribution on parameters involved in the nonlocal constitutive law, and associated modeling discrepancies relative to higher fidelity computations. As an application, we consider the propagation of stress waves through a one-dimensional heterogeneous bar with randomly generated microstructure. Several numerical tests illustrate the construction, enabling UQ in nonlocal model predictions. Although nonlocal models have become popular means for homogenization, their statistical calibration with respect to high-fidelity models has not been presented before. This work is a first step towards statistical characterization of nonlocal model discrepancy in the context of homogenization.
\end{abstract}

\renewcommand*{\thefootnote}{\arabic{footnote}}

\section{Introduction}


Nonlocal models have become viable alternatives to partial differential equation (PDE) models when the effects of the small-scale behavior of a system affect its global state \cite{beran70,cher06,karal64,rahali15,smy00,willis85,eringen1972nonlocal,bobaru2016handbook,du2020multiscale}. These models are characterized by integral operators (as opposed to differentiable operators) that embed time and length scales in their definition; as such, they are able to capture long-range effects that classical PDE models fail to describe \cite{silling2000reformulation}. These effects include the anomalous behavior often observed in coarse-grained measurements of a diffusive quantity, often referred to as anomalous or non-standard diffusion \cite{du2011mathematical,du2013nonlocal,suzuki2021fractional}. 
In the context of materials response, it is often impractical to solve equations at the small (e.g. micro) scale either because small-scale properties are not available and/or uncertain, or because the simulations are computationally infeasible. Thus, a lot of effort has been dedicated to homogenization theory with the purpose of designing large-scale models that accurately reproduce the effects of small-scale behavior \cite{zohdi2017homogenization,bensoussan2011asymptotic,weinan2003multiscale,efendiev2013generalized,junghans2008transport,kubo1966fluctuation,santosa1991dispersive,dobson2010sharp,ortiz1987method,moes1999simplified,hughes2004energy}. In this context, it is important to keep in mind that coarse-grained data do not obey the same governing laws that are valid at the small scale. Classical homogenization techniques (e.g., the use of effective quantities) are not satisfactory in representing coarse-grained behavior, creating the need for more accurate models that, while acting at a computationally feasible scale, still capture the effects of small-scale behavior. Due to their ability to capture long-range effects, nonlocal models are the best candidates for this task \cite{du2020multiscale}. Their use in the context of homogenization is not new and started with the development of homogenized models for subsurface transport \cite{suzuki2021fractional}. Here, fractional models were identified as the best means for describing super- and sub-diffusive effects, {and a Bayesian approach was developed in \cite{trillos2017bayesian}, to recover the order and the diffusion coefficient of an elliptic fractional partial differential equation}. More recently, the concept of nonlocal homogenization was successfully applied in the context of material response \cite{silling2021propagation,silling2022peridynamic}. Furthermore, with the explosion of machine learning, optimized nonlocal models were designed with the purpose of accurately reproducing observed coarse-grained behavior and predicting unseen behavior with the learnt model. We refer the reader to \cite{You2021,you2020data,you2021md,xu2021machine,xu2022subsurface,zhang2022metanor,demoraes2022} for several examples of the use of machine learning for the design of homogenized nonlocal operators and the rigorous analysis of its learning theory \cite{lu2022nonparametric,zhang2022metanor}. Although successful in providing a deterministic answer to the homogenization problem, none of these approaches characterizes the discrepancy incurred when using a nonlocal model as a coarse-grained, homogenized surrogate.

The characterization of surrogate/reduced-order modeling discrepancy is a well-known problem in the PDE literature, while it is absent in the context of nonlocal modeling. In this work, for the first time, we employ Bayesian inference for statistical estimation of nonlocal model parameters using statistical descriptions for the discrepancy from high-fidelity data. 
Bayesian methods~\cite{Laplace:1814,Casella:1990,Jaynes:2003,Robert:2004,Sivia:2006,Carlin:2011} have been used increasingly over recent decades for the probabilistic estimation of parameters of complex physical models given data on model predictions~\cite{Kennedy:2001,Higdon:2003,Oliver:2011,Cui:2016b,Hakim:2017,Huan:2018b}. The Bayesian framework provides a number of advantages in this regard, including means of (1) incorporating prior information which can be based on expert knowledge, prior/other experiments, or physical laws; (2) using arbitrary data models that can incorporate physical model complexity as well as generalized statistical models of discrepancy between data and model predictions; and (3) model comparison, selection, and averaging, with formal connections to information theory. We use Bayesian inference here, relying on an adaptive Markov chain Monte Carlo (MCMC) construction~\cite{Haario:2001}, to estimate parameters of a homogenized nonlocal model given data from a higher-fidelity PDE model. In this context, the discrepancy between model predictions and the data is due to predictive errors resulting from the necessary approximations employed in the homogenized nonlocal model construction. 

\smallskip
Specifically, we employ Bayesian calibration in the context of a one-dimensional, nonlocal wave equation that describes the propagation of stress waves through an elastic bar with a heterogeneous microstructure. We use Bayesian inference to learn the posterior probability distributions of parameters of the nonlocal constitutive law (i.e. the nonlocal kernel function). 
The data model includes an additive noise term to represent the discrepancy between the high-fidelity, PDE models and the (homogenized) nonlocal model. Our key contributions are summarized below.

\begin{enumerate}
\item We introduce and demonstrate, for the first time, a Bayesian inference framework for the statistical estimation of parameters in homogenized nonlocal models.
\item We develop a multi-step algorithm that enables efficient Bayesian inference via effective initialization and hyper-parameter choice.
\item Through several numerical tests we illustrate the prediction capability of the proposed approach by successfully using the calibrated model to predict unseen scenarios.
\end{enumerate}

\textbf{Paper outline. }
In Section ``Dispersion in Heterogeneous Materials'' we introduce the classical problem of stress wave propagation through heterogeneous materials and describe the numerical solver used for the generation of the high-fidelity data set. In Section ``Background and Related Mathematical Formulation'' we summarize relevant prior work on (deterministic) NOR: we describe the mathematical formulation of the regression problem and discuss the choice of hyperparameters for efficient training. In Section ``Bayesian Nonlocal Operator Regression'' we introduce a Bayesian approach to nonlocal model calibration. We break down the MCMC-based algorithm in several steps with the purpose of initializing distributions and hyperparameters in an efficient manner. In Section ``Application to a Heterogeneous Elastic Bar'' we illustrate through several numerical tests how our approach predicts the posterior distribution of the nonlocal model predictions, enabling uncertainty quantification (UQ) in nonlocal simulations. Section ``Conclusion'' concludes the paper with a summary of our contributions and a list of potential follow-up work.

\section{Dispersion in Heterogeneous Materials}\label{sec:DHM}

\begin{figure}[h!]
\centering
\includegraphics[width=0.5\columnwidth]{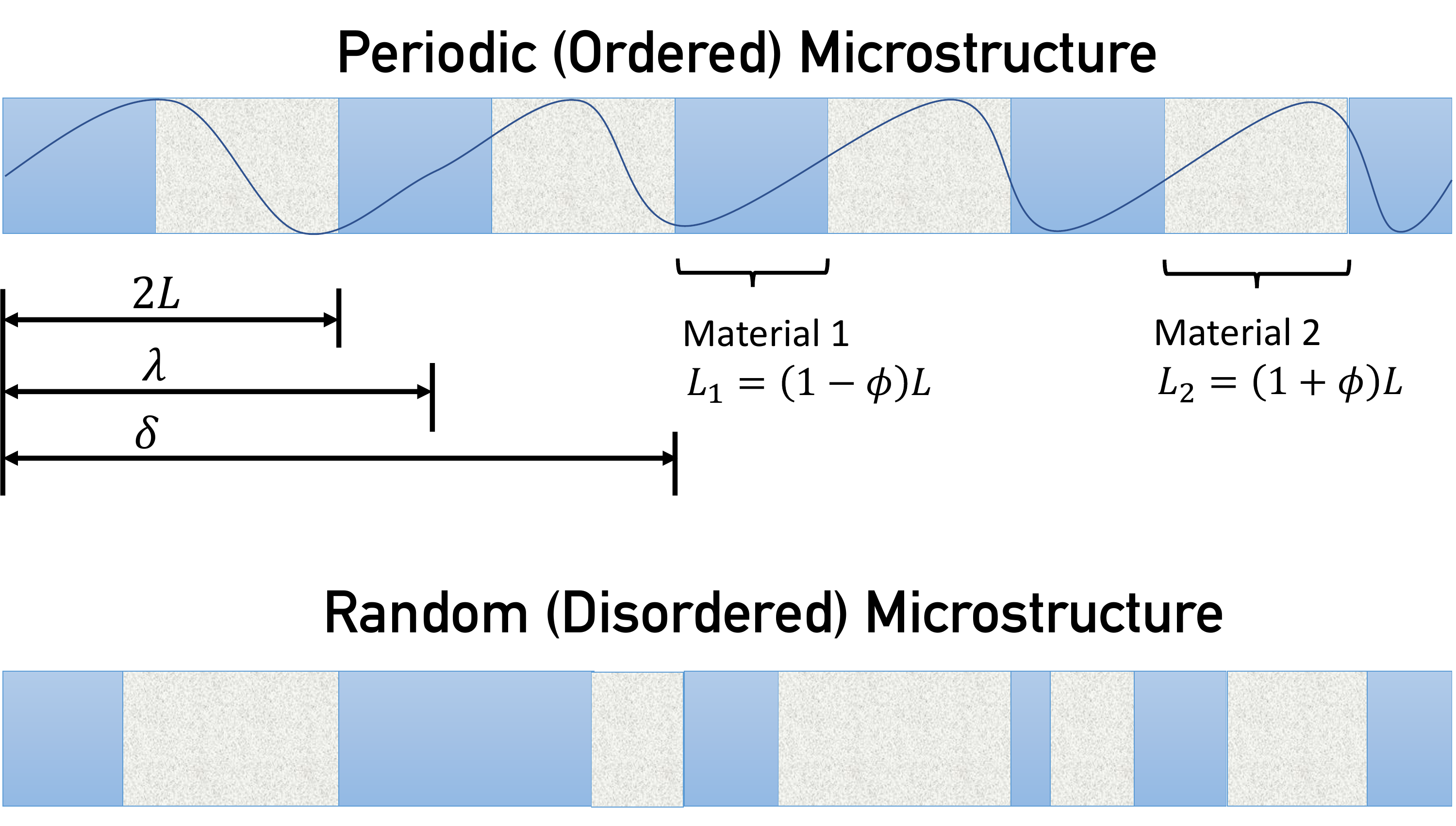}\qquad
\includegraphics[width=0.4\textwidth]{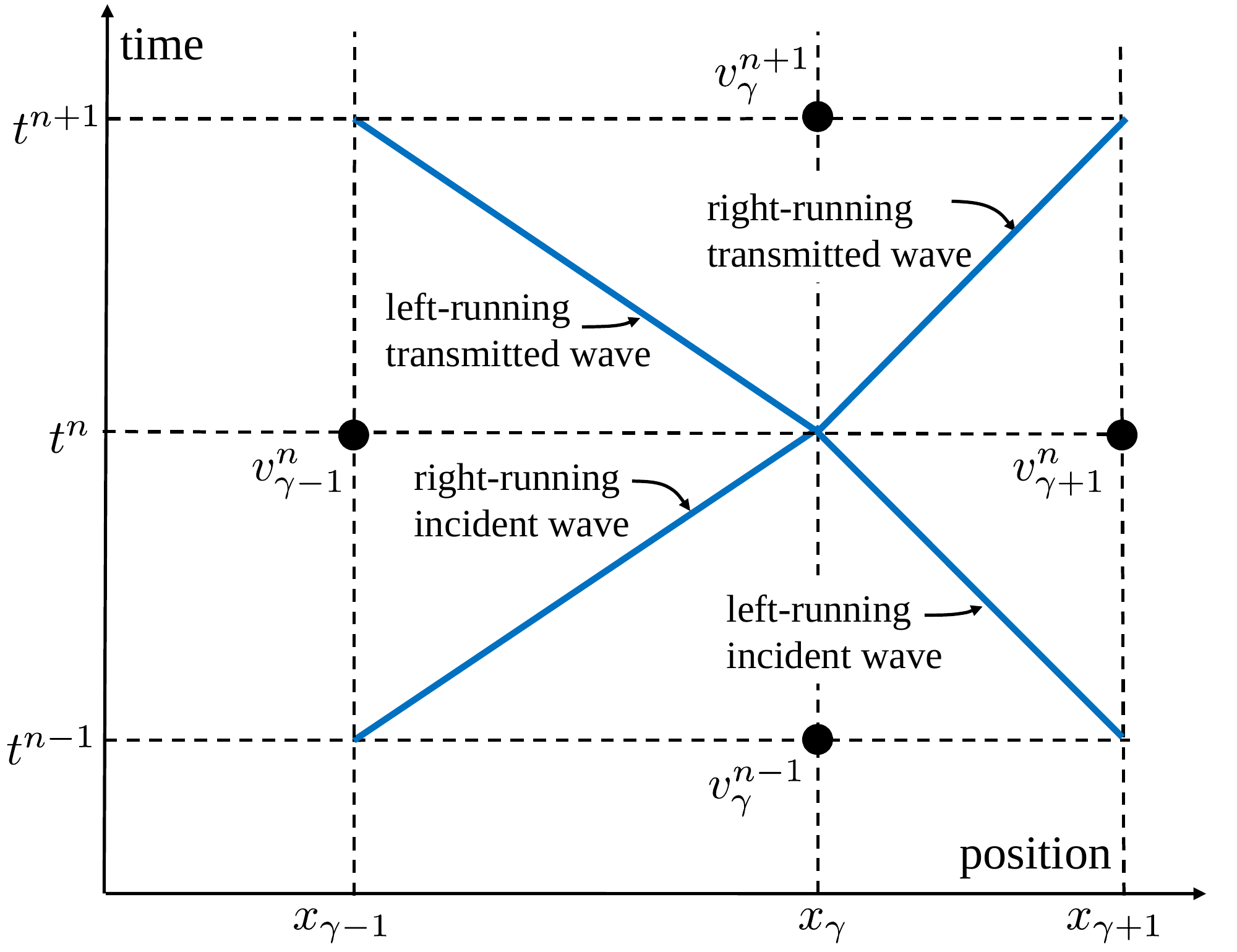}
\caption{Left: One-dimensional heterogeneous bars composed of materials 1 and 2. 
The oscillatory curve represents a moving wave with wavelength $\lambda$.
The horizon $\delta$ for the nonlocal continuum model is shown. 
Left Top: Periodic microstructure with period $2L$. Left Bottom: Random microstructure. Right: Interaction of two waves in the DNS method.}
\label{fig:bar}
\end{figure}

The classical (local) wave equation for a heterogeneous one-dimensional linearly elastic body reads as follows:
\begin{equation}
    \rho(x)\frac{\partial^2 u}{\partial t^2}(x,t)=\frac{\partial}{\partial x}\left(E(x)\frac{\partial u}{\partial x}(x,t)\right)+f(x,t)
    \label{eqn-classicalwave}
\end{equation}
where $x$ is the reference position, $\rho$ is the mass density, $t$ is time, $u$ is the displacement, 
$f$ is the external load density (assumed to be continuous), and $E$ is the elastic modulus which varies spatially.
The quantity in the large parentheses in \eqref{eqn-classicalwave} is the stress, usually denoted by $\sigma$, 
which is continuous even if $E$ has jump discontinuities.
If $E$ has a jump discontinuity at some $x_0$, the following jump conditions hold:
\begin{equation}
    [u]=0, \qquad \left[E\frac{\partial u}{\partial x}\right]=0
    \label{eqn-jumpcond}
\end{equation}
where the notation for jumps is $[w]=w(x_0^+)-w(x_0^-)$ for any function $w$.
In this paper, $E$ is piecewise continuous.
The particular choice of the function $E(x)$ describes a microstructure.
For simplicity, $E$ is assumed to take on only one of two values, but the spatial distribution 
can be either periodic or random (see the left plot of Figure~\ref{fig:bar}).
Also for simplicity, $\rho$ is assumed to be constant throughout the body.
Open subregions that have constant $E$ will be called ``grains'' in analogy with materials science.

In the case of a random microstructure, the grain sizes $L_1$ and $L_2$ for materials 1 and 2
are generated using the following expressions:
\begin{eqnarray}
  L_1&\sim& \mathcal{U}\big[(1-D)(1-\phi)L,\;(1+D)(1-\phi)L\big] \nonumber\\
  L_2&\sim& \mathcal{U}\big[(1-D)(1+\phi)L,\;(1+D)(1+\phi)L\big]
\end{eqnarray}
where $L$ is the mean value of grain size and $\phi$ determines the volume fraction of material 2. 
In this work, we take $\phi=0$ without loss of generality. This choice leads to equal volume fractions for the two materials. $\mathcal{U}[a,b]$ denotes a uniform random distribution between $a$ and $b$.
$D\in[0,1]$ is the disorder parameter that determines the overall variation in grain size,
with $D=0$ leading to a periodic microstructure with no randomness.
Within any grain, any wave $q$ of the following form is a solution to \eqref{eqn-classicalwave}:
\begin{equation}
    v(x,t)=q(x\pm ct), \qquad c=\sqrt{\frac E \rho}
    \label{eqn-dalembert}
\end{equation}
where $c$ is the wave speed and $v=\partial u/\partial t$ is the velocity of a material point.
In the expression $x \pm ct$, the $+$ sign is for a ``left-running'' wave and the $-$ is for a 
``right-running'' wave.
Because the system is linear, arbitrary superpositions of waves in both directions are also solutions to \eqref{eqn-classicalwave}.
In the direct numerical simulation (DNS) technique that is used in this study, the waves that are superposed are 
composed of finite jumps in velocity, so that $q=H$ in \eqref{eqn-dalembert}, where $H$ denotes the Heaviside step function.

\begin{figure}  [t!]
\centering
\subfigure{\includegraphics[width=.53\columnwidth]{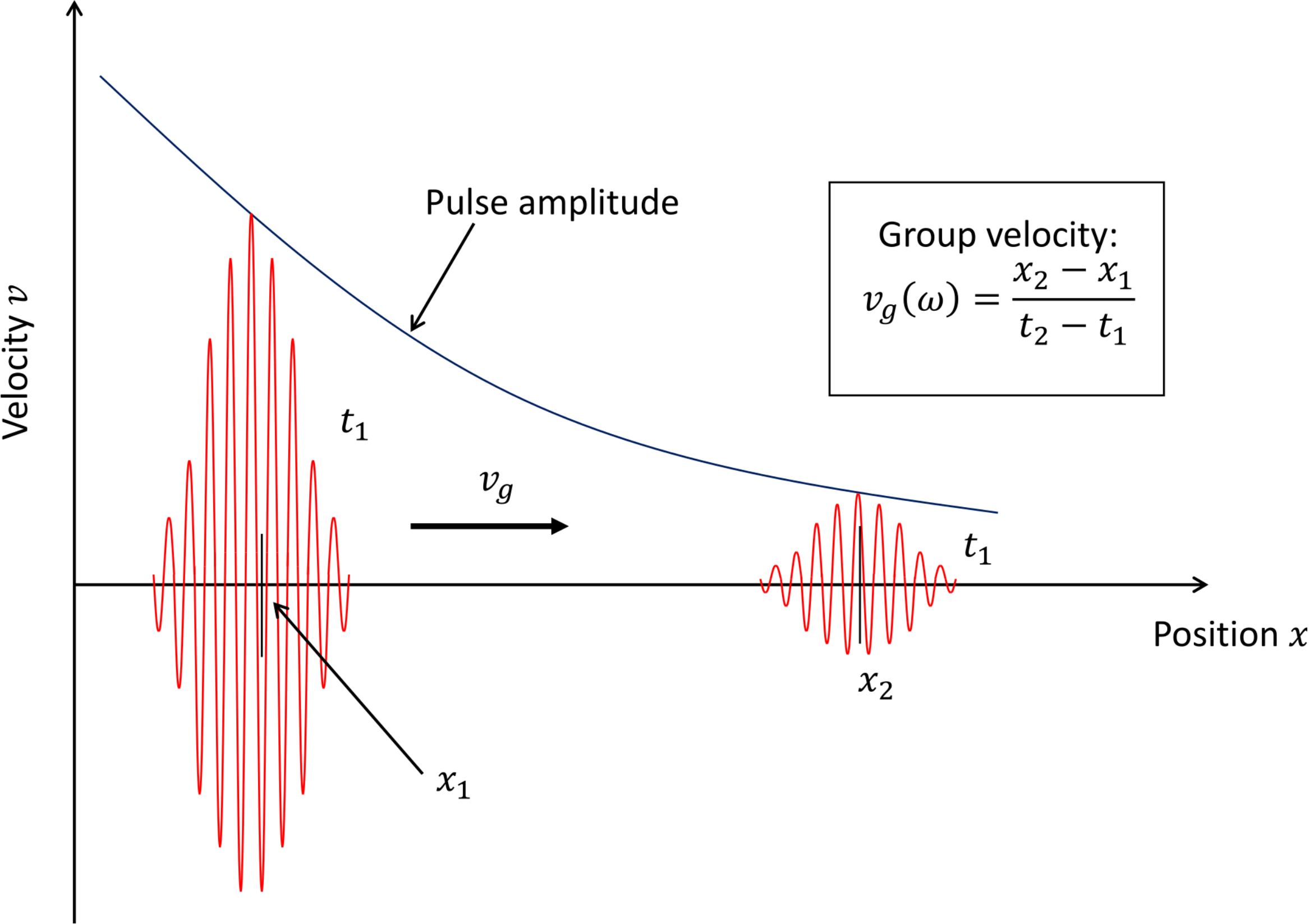}}\qquad
\subfigure{\includegraphics[width=.38\columnwidth]{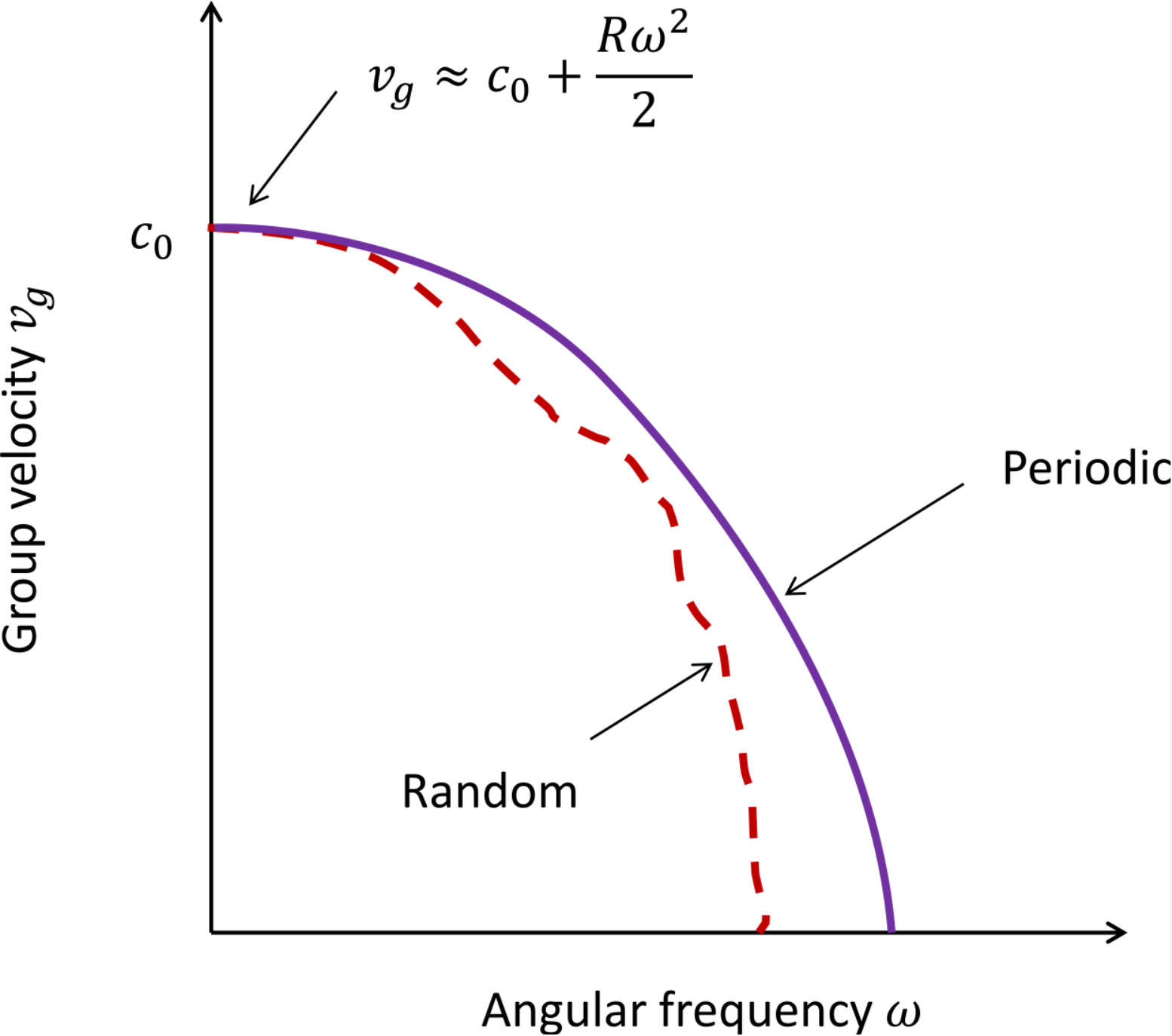}}
\caption{Left: Group velocity estimation from the speed of a wave packet, which can
attenuate significantly in random media.
Right: Qualitative dependence of group velocity on frequency.}
\label{fig-vgtypical}
\end{figure}

Later in the paper, the group velocity $v_g(\omega)$ will be used, where $\omega$
is the angular frequency of a wave.
In a homogeneous medium, the group velocity is defined by 
$v_g(\omega)=\frac{d\omega}{dk}$, 
where $k=2\pi/\lambda$ is the wave number.
With the DNS solver or any other simulation method, the group velocity can be estimated from 
the velocity of a wave packet with nominal frequency $\omega$ (see Figure~\ref{fig-vgtypical}).
In both periodic and random bars, the qualitative dependence of $v_g$ on $\omega$ is similar.
Both have the same large-wavelength (low frequency) dependence, because $\lambda\gg L$ for long
waves.
In particular, both $v_g$ functions have the same curvature as $\omega\rightarrow0$.
At smaller wavelengths, there are differences in $v_g$ between periodic and random systems.
The method for evaluating $v_g$ from wave packets is also susceptible to error near the band stop,
because in this limit, the wave packet attenuates to nearly undetectable amplitudes as it propagates.

The goal of our learning algorithm is to learn a surrogate model that is able to 
predict wave propagation over distances that are much larger than the size of the microstructure and provide error estimation. 
For both training and validation purposes, we first generate high-fidelity 
data by solving the classical wave equation within a detailed model of the 
microstructure using the DNS solver, which will now be briefly described. 
For a given forcing term $f(x,t)$ and boundary and initial conditions, 
the DNS solver provides a solution for the velocity field $v(x,t)$. 
The bar is discretized into nodes, 
$\{x_{\gamma}\}$, such that it takes a constant amount of time $\Delta t_{DNS}$ for a 
wave to travel through the cell between $x_\gamma$ and
$x_{\gamma+1}$, regardless of the elastic wave speed in the material 
between these two nodes (see the right plot of Figure~\ref{fig:bar} and \cite{You2021} for further details).
The interval between two adjacent nodes is called a ``cell.''
Each cell is composed entirely of one material or the other, with elastic modulus $E_1$ or $E_2$.
These materials have wave speeds $c_1$ and $c_2$, given by  \eqref{eqn-dalembert}.

The right plot of Figure~\ref{fig:bar} shows left-running and right-running waves on an $x$-$t$ diagram
(also called a wave diagram).
Because the wave speed varies between cells, the slopes shown in the
diagram also vary.
Because the wave transit time within any cell must be constant,
the spacing between DNS nodes varies along the bar.
The DNS solver works by assuming that in each time step, a step wave
travels from any node $\gamma$ to its neighbor on the left or right.
These neighbors also send waves to node $\gamma$.
Within any cell, the two waves obey the jump condition
\begin{equation}
   [\sigma]=\pm\rho c [v],
\label{eqn-dns-jump}
\end{equation}
which is a consequence of the momentum balance.
Note that the jump condition \eqref{eqn-dns-jump} applies to the moving waves,
while \eqref{eqn-jumpcond} applies at fixed points.
From these conditions, the DNS solver computes updated the material velocity $v_\gamma^{n+1}$ 
explicitly from the values at the adjacent nodes in time step $n$.
Details of the DNS solver can be found in \cite{silling2021propagation,You2021}.

After the velocity $v_\gamma^{n+1}$ is computed, the displacement is updated by integrating the
velocity over time:
\[    u_{DNS}(x_\gamma,t^{n+1})\approx u_\gamma^{n+1} = u_\gamma^n + \Delta t_{DNS} v_\gamma^{n+1}.   \]
The displacements are mapped onto a mesh with constant spacing by interpolation between the
irregularly spaced DNS node positions.
The mapped displacements are used as training and validation data as described below.

Since this DNS solver does not use numerical approximations for derivatives,
it does not introduce truncation error, nonphysical dispersion of waves, or similar 
undesirable numerical effects.
It allows us to model the propagation of waves through many thousands of microstructural interfaces 
and generate the high-fidelity solution for the corresponding displacement field.

\section{Background and Related Mathematical Formulation}\label{sec:background}

In this section we introduce the nonlocal surrogate model used in this work and the proposed Bayesian learning approach. Consider the simulation on a spatial domain $\Omega$ and time domain $[0,T]$, and $S$ observations of forcing terms $f^s(x,t)$ and their corresponding high-fidelity solution and/or experimental measurements of displacement fields $u_{DNS}^s(x,t)$. In the following we employ these solutions as the ground-truth dataset, denoted as  $\mcD:=\{u^s_{DNS}\}_{s=1}^S$. Here we assume that both $f^s$ and $u^s_{DNS}$ measurements are provided on a set of time instances $t^n\in[0,T]$ and discretization points $x_i\in\Omega$. Without loss of generality, we assume that all measurements are provided on uniformly spaced spatial and time instances, with fixed spatial grid size $\Delta x$ and time step size $\Delta t$, and we denote the collection of all discretization points as $\chi=\{x_i\}_{i=1}^L$. The overall goal of nonlocal operator regression is to provide a nonlocal model surrogate for the simulation of wave propagation in heterogeneous materials. Thus, we claim that given the same forcing terms $f^s(x,t)$, the corresponding solution, $u_{NL}^s(x,t)$, $(x,t)\in\Omega\times[0,T]$, of the nonlocal surrogate model provides a good approximation of the ground-truth data, i.e., $u_{NL}^s(x,t)\approx u_{DNS}^s(x,t)$.

Throughout this paper, for any vector $\vb=[v_1,\cdots,v_q]\in\real^q$, we use $\vertii{\vb}_{l^2}$ to denote its $l^2$ norm, i.e., $\vertii{\vb}_{l^2}:=\sqrt{\sum_{i=1}^qv_i^2}$. 
For a function $u(x,t)$ with $(x,t)\in\Omega\times[0,T]$, its discrete $l^2$ norm is defined as
$$\vertii{u}_{l^2(\Omega\times[0,T])}:=\sqrt{\Delta t\Delta x\sum_{n=0}^{T/\Delta t} \sum_{x_i\in\chi} u^2(x_i,t^n)},$$
which can be interpreted as a numerical approximation of the $L^2(\Omega\times[0,T])$ norm of $u$. Finally, in what follows, $\mathbf{I}_p$ denotes the $p\times p$ identity matrix; with an abuse of notation, when there is no confusion on its dimension, we simply use $\mathbf{I}$ to denote the identity matrix.

\subsection{Nonlocal Operator Regression: learning nonlocal kernels}\label{sec:nor}

In this section we review the general nonlocal operator regression (NOR) approach developed in \cite{You2021}.
The goal of NOR is to find a nonlocal model that best describes the evolution of a homogenized quantity such as the propagation of stress waves in highly heterogeneous materials. 

NOR starts from the assumption that a high-fidelity data set, satisfying an underlying high-fidelity model, is available. Given a force loading term $f(x,t)$, $(x,t)\in\Omega\times[0,T]$, proper boundary conditions and initial conditions, we represent the high-fidelity (HF) model as:
\begin{equation}
\dfrac{\partial^2 {u}_{HF}}{\partial t^2}(x,t)-\mcL_{HF}[{u}_{HF}](x,t)=f(x,t).   
\end{equation}
Here, $\mcL_{HF}$ is the HF operator that accounts for the detailed microstructure, and ${u}_{HF}(x,t)$ is the HF solution which can be provided either from fine-scale simulations or from experimental measurements in practice. In this work, we take $u_{HF}$ as the numerical solution generated from the DNS solver as described in Section ``Dispersion in Heterogeneous Materials'', i.e., $u_{HF}=u_{DNS}$. Analogously, we will refer to the homogenized effective nonlocal (NL) model as the homogenized surrogate, and assume it has the form
\begin{equation}\label{eqn:homo}
\dfrac{\partial^2 {u}_{NL}}{\partial t^2}(x,t)-\mcL_{NL}[{u}_{NL}](x,t)=f(x,t),   
\end{equation}
where the operator $\mcL_{NL}:=\mcL_{K_\Cb}$ is an integral operator associated with a nonlocal kernel $K_{\Cb}$: 
\begin{equation}\label{eqn:intoperator}
    \mcL_{K_\Cb}[u](x,t)=\int_{\omg} K_{\Cb}(x,y)(u(y,t)-u(x,t))dy,
\end{equation}
and ${u}_{NL}(x,t)$ is the nonlocal solution. Here $\Cb:=\{C_m\}_{m=0}^M\in\real^{M+1}$ is the parameter set of kernel {$K_\Cb$}, which will be optimized during training. As shown in \cite{du2017peridynamic}, the second-order-in-time nonlocal equation in \eqref{eqn:homo} is guaranteed to be well-posed as far as the kernel $K_\Cb$ is uniformly Lipschitz continuous. That means, the resultant surrogate model is guaranteed to be solvable in applications, when proper boundary conditions and numerical discretization methods are employed. Following \cite{you2020data,you2021md,You2021}, we choose to take $K_\Cb$ as a radial, sign-changing, nonlocal kernel function, compactly supported on the ball of radius $\delta$ centered at $x$, i.e., $B_\delta(x)$. Then, we parameterize the nonlocal kernel $K_\Cb$ as a linear combination of Bernstein basis polynomials: 
\begin{equation}\label{model-kernel}
\begin{aligned}
K_\Cb\left(\frac{|\xi|}{\delta}\right) 
& = \sum_{m=0}^{M}\frac{C_m}{\delta^{d+2}} B_{m,M}
    \bigg (\bigg|\frac{\xi}{\delta}\bigg|\bigg),
\end{aligned}
\end{equation}
where $d=1$ is the dimension of the physical domain, $\Omega$, and the Bernstein basis functions are defined as
$$B_{m,M}(\xi) = \begin{pmatrix}
M\\
m\\
\end{pmatrix}
\xi^m(1-\xi)^{M-m}\;\; \text{ for } 0\leq \xi\leq 1.$$
This choice ensures that the learnt model can be readily applied in simulation problems.

To obtain the optimal nonlocal surrogate operator $\mcL_{NL}$, the nonlocal operator regression approach seeks the best $\Cb$ through an optimization-based procedure. For each forcing term $f^s(x,t)$, let $u^s_{NL,\Cb}(x,t)$, $(x,t)\in \Omega\times [0,T]$ be the nonlocal solution of \eqref{eqn:homo} corresponding to a specific set of kernel parameter $\Cb:=\{C_m\}_{m=0}^{M}$. We aim to find the optimal parameter set $\Cb$ such that the (approximated) nonlocal solution $u^s_{NL,\Cb}(x_i,t^n)$ for a common loading $f^s$ is as close as possible to the HF solution $u^s_{DNS}(x_i,t^n)$, for all provided observation pairs $\{(u^s_{DNS},f^s)\}_{s=1}^S$ and $x_i\in\chi$, $t^n=0,\cdots,T/\Delta t$. To do so, for each given forcing term $f^s(x,t)$ we need to provide a numerical approximation of its corresponding nonlocal solution, $(u^s_{NL,\Cb})_{i}^n\approx u^s_{NL,\Cb}(x_i,t^n)$. Here, with the initial condition $(u^s_{NL,\Cb})_{i}^0:=u^s_{DNS}(x_i,0)$ and proper boundary conditions, we discretize \eqref{eqn:homo} with the central difference scheme in time and Riemann sum approximation of the nonlocal operator in space, and obtain the approximated nonlocal solution at time $t^{n+1}$ as follows: 
\begin{align}
\nonumber(u^s_{NL,\Cb})_{i}^{n+1}:=& 2(u^s_{NL,\Cb})_{i}^n-(u^s_{NL,\Cb})_{i}^{n-1}+\Delta t^2f^s(x_i,t^n)+\Delta t^2 \left(\mcL_{K_\Cb,h}[u^s_{NL,\Cb}]\right)_i^n\\
\nonumber=&2(u^s_{NL,\Cb})_{i}^n-(u^s_{NL,\Cb})_{i}^{n-1}+\Delta t^2f^s(x_i,t^n)\\
&+\Delta t^2\Delta x\sum_{x_j\in B_\delta(x_i)\cap\chi} K_\Cb(|x_j-x_i|)((u^s_{NL,\Cb})_{j}^n-(u^s_{NL,\Cb})_{i}^n),\label{eqn:tildeu}
\end{align}
where $\mcL_{K_\Cb,h}$ is an approximation of $\mcL_{K_\Cb}$ by the Riemann sum with uniform grid spacing $\Delta x$.
The optimal parameters $\Cb^*=\{C^*_m\}$ and the corresponding nonlocal surrogate operator $\mcL_{NL}:=\mcL_{K_{\Cb^*}}$ can be obtained by considering the normalized squared-loss of displacement in $\omg\times[0,T]$:
\begin{align}
\Cb^*=\underset{\Cb}{\text{argmin}}&\sum_{s=1}^S\dfrac{\vertii{u_{NL,\Cb}^s-u_{DNS}^s}_{l^2(\Omega\times[0,T])}^2}{\vertii{u_{DNS}^s}_{l^2(\Omega\times[0,T])}^2}+\lambda \vertii{\Cb}_{l^2}^2,\label{eqn:opt}\\
\text{s.t. }&\mcL_{K_{\Cb}} \text{ satisfies physics-based constraints.}\label{eqn:optcond2}
\end{align}
Here $\lambda$ is a regularization parameter, and \eqref{eqn:optcond2} depends on the partial physical knowledge of the heterogeneous material, which we will discuss later on in Section ``Physics Constraints''.

\subsection{Initialization for Parameters and Hyper-parameters}\label{sec:lcurve}

We note that the solution of \eqref{eqn:tildeu} has numerical errors accumulated from $t=0$ to $t=T$, and hence in practice the optimization problem \eqref{eqn:opt} aims to minimize the accumulated error of displacement fields. This setting was originally proposed and employed in \cite{You2021}, where the authors found that minimizing the accumulated error would help to learn physically-stable surrogate models, and such a stability plays a critical role in long-term prediction tasks. On the other hand, we also point out that in some application scenarios one can also choose to minimize the step-by-step time integration error. In \cite{lu2022nonparametric,zhang2022metanor}, an approximated nonlocal solution with one-step temporal error was considered:
\begin{align*}
\nonumber&(\tilde{u}^s_{NL,\Cb})_{i}^{n+1}:= 2(u^s_{DNS})_{i}^n-(u^s_{DNS})_{i}^{n-1}+\Delta t^2f^s(x_i,t^n)+\Delta t^2 \left(\mcL_{K_\Cb,h}[u^s_{DNS}]\right)_i^n\\
&=2(u^s_{DNS})_{i}^n-(u^s_{DNS})_{i}^{n-1}+\Delta t^2f^s(x_i,t^n)+\Delta t^2\Delta x\sum_{x_j\in B_\delta(x_i)\cap\chi} K_\Cb(|x_j-x_i|)((u^s_{DNS})_{j}^n-(u^s_{DNS})_{i}^n),
\end{align*}
where $(u^s_{DNS})_{i}^n:=u^s_{DNS}(x_i,t^n)$ represents the high-fidelity solution. A step-by-step loss of displacement is then formulated as
\begin{align}
\tilde{\Cb}^*=\underset{\Cb}{\text{argmin}}&\sum_{s=1}^S\dfrac{\vertii{\tilde{u}_{NL,\Cb}^s-u_{DNS}^s}_{l^2(\Omega\times[0,T])}^2}{\vertii{u_{DNS}^s}_{l^2(\Omega\times[0,T])}^2}+\lambda \vertii{\Cb}_{l^2}^2,\label{eqn:1stepopt}\\
\text{s.t. }&\mcL_{K_{\Cb}} \text{ satisfies physics-based constraints.}\label{eqn:1stepoptcond}
\end{align}
The loss function in \eqref{eqn:1stepopt} is in fact a quadratic equation with respect to $\Cb$. As such, NOR is equivalent to a linear regression model and solving \eqref{eqn:1stepopt} becomes a trivially linear problem. This fact makes hyper-parameter tuning, such as the selection of the regularization parameter $\lambda$, more efficient in \eqref{eqn:1stepopt}. 
For instance, one can identify the optimal $\lambda$ as the maximizer of the curvature of the curve, following the L-curve method \cite{hansen_LcurveIts_a,LangLu22,lu2022nonparametric}. 
Let $l$ be a parametrized curve in $\R^2$, satisfying
\begin{equation}\label{eqn:lcurve}
l(\lambda) = (\alpha(\lambda), \beta(\lambda)) := (\text{log}(\mathcal{E}(\Cb^*{(\lambda)})), \text{log}(\mathcal{R}(\Cb^*{(\lambda)})),
\end{equation}
where $\mathcal{E}(\Cb^*{(\lambda)}) := \sum_{s=1}^S\frac{\vertii{\tilde{u}_{NL,\Cb^*{(\lambda)}}^s-u_{DNS}^s}_{l^2(\Omega\times[0,T])}^2}{\vertii{u_{DNS}^s}_{l^2(\Omega\times[0,T])}^2}$ is the loss function without regularization, and $ \mathcal{R}(\Cb^*{(\lambda)})$ is the regularization term, which is taken as $\vertii{\Cb^*{(\lambda)}}^2_{l^2}$ in \eqref{eqn:1stepopt}. Here, we used $\Cb^*{(\lambda)}$ to highlight the fact that the optimal parameter, $\Cb^*$, depends on the choice of the regularization parameter $\lambda$. The optimal parameter for $\lambda$ then taken as the maximizer of the curvature of $l$ as 
\begin{align}\label{eq:opt_lambda}
\smash{	\lambda^*
	= \underset{\lambda}{\text{argmax}} \frac{\alpha'\beta'' - \alpha' \beta''}{(\alpha'\,^2 + \beta'\,^2)^{3/2}}.
}\end{align}
This optimal parameter $\lambda^*$ balances the loss $\mathcal{E}$ and the regularization. For further details of the L-curve method and discussions, we refer interested readers to \cite{hansen_LcurveIts_a}.

In this work, we use the step-by-step loss formulation in \eqref{eqn:1stepopt} to provide an efficient estimation for $\Cb$ and to select the optimal regularization parameter $\lambda$. Then, the selected parameter set and hyperparameter will be employed in the optimization problem with accumulated loss, as an initial guess for the optimization solver and an estimated regularization parameter, respectively. Further details and discussions will be provided in Section ``A Two-Phase Learning Algorithm''.

\subsection{Physics Constraints}\label{sec:physics}

As illustrated in \cite{You2021}, when some physical knowledge is available, such as the effective wave speed for infinitely long wavelengths and the curvature of the dispersion curve in the low-frequency limit, this knowledge can be incorporated into the optimization problem as physics-based constraints in \eqref{eqn:optcond2} and \eqref{eqn:1stepoptcond}. In particular, when the effective wave speed for infinitely long wavelengths, $c_0$, is available, the corresponding physics-based constraint is:
\begin{equation}\label{eqn:cond1}
    \int_{0}^\delta \xi^2K_\Cb(|\xi|)d\xi=\rho c_0^2,
\end{equation}
where $\rho$ is the effective material density. Discretizing \eqref{eqn:cond1} by Riemann sum, we obtain the first constraint on $\{C_m\}$:
\begin{equation}\label{eqn:cond1_disc}
    \rho c_0^2=\sum_{m=0}^{M} C_m\sum_{\eta=1}^{\lfloor\delta/\Delta x\rfloor} \dfrac{\eta^2 \Delta x^3}{\delta^3} B_{m,M}
    \bigg (\bigg|\frac{\eta\Delta x}{\delta}\bigg|\bigg)=\sum_{m=0}^{M} C_m A_{1m}
\end{equation}
where $A_{1m}:=\sum_{\eta=1}^{\lfloor\delta/\Delta x\rfloor} \frac{\eta^2 \Delta x^3}{\delta^3} B_{m,M}\bigg (\bigg|\frac{\eta\Delta x}{\delta}\bigg|\bigg)$. Furthermore, when the curvature of the dispersion curve in the low-frequency limit, $R$, is also available, the corresponding physics-based constraint is:
\begin{equation}\label{eqn:cond2}
    \int_{0}^\delta \xi^4 K_\Cb(|\xi|)d\xi=-4\rho c_0^3R.
\end{equation}
Discretizing \eqref{eqn:cond2} with the Riemann sum approximation yields the second constraint on $\{C_m\}$:
\begin{equation}\label{eqn:cond2_disc}
    -4\rho c_0^3 R=\sum_{m=0}^{M} C_m\sum_{\eta=1}^{\lfloor\delta/\Delta x\rfloor} \dfrac{\eta^4 \Delta x^5}{\delta^3} B_{m,M}
    \bigg (\bigg|\frac{\eta\Delta x}{\delta}\bigg|\bigg) =\sum_{m=0}^{M} C_m A_{2m}
\end{equation}
where $A_{2m}:=\sum_{\eta=1}^{\lfloor\delta/\Delta x\rfloor} \frac{\eta^4 \Delta x^5}{\delta^3} B_{m,M}\bigg (\bigg|\frac{\eta\Delta x}{\delta}\bigg|\bigg)$. These two physics-based equations are imposed as linear constraints on $\{C_m\}$. In this work, we consider the heterogeneous bar composed of alternating layers of two dissimilar materials, with (averaged) layer sizes $L_1=(1-\phi)L$, $L_2=(1+\phi)L$ for materials 1 and 2, respectively. Then the effective material density, Young's modulus and the wave speed are given by $\rho=((1-\phi)\rho_1+(1+\phi)\rho_2)/2$, $E=2/((1-\phi)E_1^{-1}+(1+\phi)E_2^{-1})$, and $c_0=\sqrt{E/\rho}$. 
$R$ is the second derivative of the wave group velocity with respect to frequency at $\omega$=0. 
As discussed in Section ``Dispersion in Heterogeneous Materials'', the group velocity is estimated from the speed of wave packets using the DNS solver.
For the periodic microstructure, the curvature $R$ of the function $v_g(\omega)$ at $\omega=0$ is found by numerically differentiating
this function.
The value obtained for the present bar composition is
$R=\dfrac{d^2v_g^{DNS}}{d\omega^2}(0)=-0.006135$. 
The same value of $R$ is also employed for the random microstructure because, as discussed in Section ``Dispersion in Heterogeneous Materials'', the same
dispersion properties apply to both periodic and random media with the same $\phi$ and $L$
if the wavelength is much greater than the microstructural length scale.

When applying \eqref{eqn:cond1_disc} and \eqref{eqn:cond2_disc} in the optimization problem \eqref{eqn:opt} or \eqref{eqn:1stepopt}, we reformulate this constrained optimization problem such that an unconstrained optimization problem is obtained. In particular, denoting $\Cb_L:=[C_0,\cdots,C_{M-2}]^T$ and $\Cb_R:=[C_{M-1},C_{M}]^T$, \eqref{eqn:cond1_disc} and \eqref{eqn:cond2_disc} can be rewritten as
\begin{equation*}
\mathbf{A}_L\Cb_L+\mathbf{A}_R\Cb_R=
\left[\begin{array}{c}
\rho c_0^2\\
-4\rho c_0^3R\\
\end{array}\right], \text{ where }\mathbf{A}_L:=\left[\begin{array}{ccc}
A_{10}&\cdots&A_{1,M-2}\\
A_{20}&\cdots&A_{2,M-2}\\
\end{array}\right], \; \mathbf{A}_R:=\left[\begin{array}{cc}
A_{1,M-1}&A_{1,M}\\
A_{2,M-1}&A_{2,M}\\
\end{array}\right].
\end{equation*}
Hence, one can eliminate $\Cb_R$ by writting it as a linear expression in $\Cb_L$:
\begin{equation}
\Cb_R=\mathbf{A}^{-1}_R \left(
\left[\begin{array}{c}
\rho c_0^2\\
-4\rho c_0^3R\\
\end{array}\right]-\mathbf{A}_L\Cb_L\right),  
\end{equation}
and substituting into the optimization problem \eqref{eqn:opt} or \eqref{eqn:1stepopt}. Therefore, in the later sections we only need to solve for $\Cb_L$, and will demonstrate the algorithm for the unconstrained problem.

\section{Bayesian Nonlocal Operator Regression}\label{sec:bnor}

Given observations of forcing terms $f^s(x,t)$, the corresponding DNS solution of displacement fields at time instance $t^n\in[0,T]$, and discretization points $x_i\in\chi$, in this section we formulate the Bayesian inference problem. Here, we stress that while the DNS data can be observed, the ground truth model is unknown. Moreover, due to the fact that the target surrogate homogenized model is supposed to act at a larger scale without resolving the detailed microstructure, the model discrepancy would be the largest contributor to the overall modeling error and correspondingly the predictive uncertainty. 
For the $s-$th observation, we model the discrepancy between $u^s_{NL,\Cb}(x_i,t^n)$ and the ground truth measurement $u^s_{DNS}(x_i,t^n)$ 
as additive independent unbiased Gaussian random noise $\epsilon$, with: 
\begin{equation}\label{eqn:adderror}
u^s_{DNS}(x_i,t^n) = u^s_{NL,\Cb}(x_i,t^n) + \epsilon_{s,i,n},\;\qquad \epsilon_{s,i,n}\sim \mcN(0,\tilde{\sigma}_s^2).    
\end{equation}
Here, with $\sigma$ a constant independent of $s$, $\tilde{\sigma}_s:=\sigma \cdot \vertii{u^s_{NL,\Cb}(x,t)}_{l^2(\Omega\times[0,T])}$, and $\mcN(0,\tilde{\sigma}_s^2)$ represents the normal distribution with zero mean and standard deviation $\tilde{\sigma}_s$. 
Thus, $\sigma$ is the standard deviation of $\left\{u_{DNS}^s-u_{NL,\Cb}^s\right\}_{s=1}^S$ after normalization w.r.t. the $l^2$-norm of $\left\{u_{NL,\Cb}^s\right\}_{s=1}^S$. For each observation, the likelihood function at $(x_i,t^n)$ is
\begin{equation*}
    f(u_{DNS}^s(x_i,t^n)|\Cb)=\dfrac{1}{\sqrt{2\pi\tilde{\sigma}_s^2}}\exp\left(-\dfrac{\verti{u_{NL,\Cb}^s(x_i,t^n)-u_{DNS}^s(x_i,t^n)}^2}{2\tilde{\sigma}_s^2}\right).
\end{equation*}
Therefore, given the independent noise construction, the likelihood for all the observations is
\begin{equation}\label{eqn:likelihood}
    f(\mcD|\Cb)=\prod_{s,i,n=1}^{S,L,T/\Delta t}\left\{\dfrac{1}{\sqrt{2\pi\sigma^2\vertii{u_{NL,\Cb}^s}_{l^2(\Omega\times[0,T])}^2}}\exp\left({-\dfrac{\verti{u_{NL,\Cb}^s(x_i,t^n)-u_{DNS}^s(x_i,t^n)}^2}{2\sigma^2\vertii{u_{NL,\Cb}^s}_{l^2(\Omega\times[0,T])}^2}}\right)\right\}.
\end{equation}

We employ priors as a means of regularization. We use a multivariate normal prior distribution on the kernel parameter, $\Cb\sim \mcN(\hat{\Cb},\frac{\hat{\sigma}^2}{\lambda }\mathbf{I})$. Here ${\hat{\Cb}}$ is the learnt parameter from the deterministic nonlocal operator regression technique introduced in Section ``Background and Related Mathematical Formulation'', $\hat{\sigma}$ is defined as the $\sigma_2$ estimated in Step 1b of Algorithm~\ref{algorithm:1} below, 
and $\lambda$ is the chosen regularization parameter from the L-curve method, to achieve a good balance between the prior and the likelihood contributions. 
Since the normalized standard deviation of model discrepancy, $\sigma$, is anticipated to be smaller than $1$, in our implementation we infer $\log(\sigma)$ instead of $\sigma$, and limit its range to $[-10,0]$. Moreover, since we have no other prior knowledge about $\log(\sigma)$, we assume that its prior satisfies a uniform random distribution, i.e., $\log(\sigma)\sim\mathcal{U}[-10,0]$.
Then, combining the prior and likelihood in \eqref{eqn:likelihood}, 
we define the (unnormalized) posterior $\pi(\Cb|\mcD)\propto f(\mcD|\Cb)P(\Cb)$, and obtain the associated negative log-posterior formulation after eliminating the constant terms:
\begin{equation}
    -log(\pi(\Cb|\mcD))=\sum_{s=1}^S\left\{\dfrac{\vertii{u_{NL,\Cb}^s-u_{DNS}^s}_{{l^2(\Omega\times[0,T])}}^2}{2\sigma^2\vertii{u_{NL,\Cb}^s}_{{l^2(\Omega\times[0,T])}}^2}+N\log\left(\sigma\vertii{u_{NL,\Cb}^s}_{{l^2(\Omega\times[0,T])}}\right)\right\}+\lambda \dfrac{\vertii{\Cb-\hat{\Cb}}_{l^2}^2}{2{\hat{\sigma}}^2},
\label{eq:log-post}
\end{equation}
where $N=L \frac{T}{\Delta t}$ is the total number of measurements for each observation pair, $(f^s,u^s_{DNS})$. 

\subsection{A Two-Phase Learning Algorithm}\label{sec:algorithm}

\begin{algorithm}
\caption{A Two-Phase Learning Algorithm}
{{\begin{algorithmic}[1]
\State{{\bf Find a good initial state}

${\bf 1a)}$ 
Learning $\Cb_1$ by minimizing the step-by-step error via:
\begin{equation}
\Cb_1:=\underset{\Cb}{\text{argmin}}\sum_{s=1}^S\dfrac{\vertii{\tilde{u}_{NL,\Cb}^s-u_{DNS}^s}_{l^2(\Omega\times[0,T])}^2}{\vertii{u_{DNS}^s}_{l^2(\Omega\times[0,T])}^2}+\lambda\vertii{\Cb}_{l^2}^2.
\end{equation}
Then calculate $\sigma_1$ as the standard deviation of $u_{NL,\Cb_1}-u_{DNS}$, and update the regularization parameter, $\lambda_1$, following \eqref{eqn:lambda_1}.

${\bf 1b)}$ 
With initial values $\Cb_1$ and regularization parameter $\lambda_1$ obtained from step ${\bf 1a)}$, find $\Cb_2$ and $\sigma_2$ as follows: 
\begin{equation}
(\Cb_2,\sigma_2):=\underset{\Cb,\sigma}{\text{argmin}}\sum_{s=1}^S\left\{\dfrac{\vertii{u_{NL,\Cb}^s-u_{DNS}^s}_{l^2(\Omega\times[0,T])}^2}{2\sigma^2\vertii{u_{DNS}^s}_{l^2(\Omega\times[0,T])}^2}+N\log\left(\sigma\vertii{u_{DNS}^s}_{l^2(\Omega\times[0,T])}\right)\right\}+\lambda_1 \dfrac{\vertii{\Cb}_{l^2}^2}{2\sigma_1^2}.
\end{equation}

${\bf 1c)}$ With the initial values $\hat{\Cb}:=\Cb_2$, determine the regularization parameter by solving the linear regression problem with the quadratic loss function:
\begin{equation}
    \Cb_3:=\underset{\Cb}{\text{argmin}}\sum_{s=1}^S\dfrac{\vertii{\tilde{u}_{NL,\Cb}^s-u_{DNS}^s}_{{l^2(\Omega\times[0,T])}}^2}{\vertii{u_{NL,\Cb}^s}_{{l^2(\Omega\times[0,T])}}^2}+\lambda \vertii{\Cb-\hat{\Cb}}_{l^2}^2,
\end{equation}
and update the regularization parameter $\lambda_2$ following \eqref{eqn:lambda_2}.

}
\State{{\bf Perform MCMC}

With the initial values $(\Cb_2,\sigma_2)$ and the regularization parameter $\lambda_2$ from phase 1, run MCMC and sample the posterior $p(\Cb,\sigma|\mcD)$, where 
the negative log-posterior, up to an additive constant, is given by
\begin{equation}\label{eqn:post_CC0_step4a}
    \sum_{s=1}^S\left\{\dfrac{\vertii{u_{NL,\Cb}^s-u_{DNS}^s}_{l^2(\Omega\times[0,T])}^2}{2\sigma^2\vertii{u_{NL,\Cb}^s}_{l^2(\Omega\times[0,T])}^2}+N\log\left(\sigma\vertii{u_{NL}^s}_{l^2(\Omega\times[0,T])}\right)\right\}+\lambda_2\dfrac{\vertii{\Cb-\Cb_2}_{l^2}^2}{2\sigma_2^2}.
\end{equation}

\State{{\bf Postprocesing}}

Extract effective samples from the MCMC chain, and estimate statistical moments of the corresponding solutions and other quantities of interests.
}
\end{algorithmic}}}
\label{algorithm:1}
\end{algorithm}

To learn the posterior distribution \eqref{eq:log-post}, we employ an adaptive Markov chain Monte Carlo (MCMC) method \cite{Haario:2001,andrieu2003introduction,andrieu2008tutorial,DebusschereUQTk:2017}. MCMC is an effective applied Bayesian estimation procedure, generating random samples from the target posterior distribution. Particularly in high dimensional chains, MCMC benefits from a good initial guess to reduce burn-in. In order to accelerate convergence and obtain a good sampling of the posterior, we provide a good initial parameter estimate to start the chain.

To this end, a two-phase learning algorithm is proposed, with the main steps summarized in Algorithm \ref{algorithm:1}. In this algorithm, an initialization phase, denoted as phase 1, is proposed before the MCMC algorithm in phase 2, with the purpose of providing good initial values that are close enough to the \emph{maximum a posteriori} (MAP) parameter estimate from \eqref{eq:log-post}.

Phase 1 is composed of a sequence of deterministic optimization problems. First, in step \textbf{1a)} we consider the quadratic problem of minimizing the step-by-step error
\begin{align*}
\Cb_1=\underset{\Cb}{\text{argmin}}&\sum_{s=1}^S\dfrac{\vertii{\tilde{u}_{NL,\Cb}^s-u_{DNS}^s}_{l^2(\Omega\times[0,T])}^2}{\vertii{u_{DNS}^s}_{l^2(\Omega\times[0,T])}^2}+\lambda \vertii{\Cb}_{l^2}^2,\\
\text{s.t. }&\mcL_{K_{\Cb}} \text{ satisfies physics-based constraints,}
\end{align*}
to provide estimates for the parameter solution, $\Cb_1$. Based on the estimated parameter, the approximated nonlocal solution $u^s_{NL,\Cb_1}$ can then be obtained following the numerical scheme \eqref{eqn:tildeu}, and an estimate of the parameter $\sigma$, which is denoted as $\sigma_1$, is provided by calculating the (normalized) standard deviation of $u_{NL,\Cb_1}-u_{DNS}$. We also estimate the regularization parameter, $\lambda_1$, as:
\begin{equation}\label{eqn:lambda_1}
\lambda_1:=\sum_{s=1}^S\dfrac{\vertii{u_{NL,\Cb_1}^s-u_{DNS}^s}_{l^2(\Omega\times[0,T])}^2}{\vertii{u_{DNS}^s}_{l^2(\Omega\times[0,T])}^2\vertii{\Cb_1}_{l^2}^2}.   
\end{equation}
Intuitively, this parameter guarantees that the accumulated error and the regularization term have similar scales. Then, based on the estimated parameter $\Cb_1$ and the regularization parameter $\lambda_1$, in step \textbf{1b)} we solve a deterministic optimization problem:
\begin{align}
(\Cb_2,\sigma_2):=\underset{\Cb,\sigma}{\text{argmin}}&\sum_{s=1}^S\left\{\dfrac{\vertii{u_{NL,\Cb}^s-u_{DNS}^s}_{l^2(\Omega\times[0,T])}^2}{2\sigma^2\vertii{u_{DNS}^s}_{l^2(\Omega\times[0,T])}^2}+N\log\left(\sigma\vertii{u_{DNS}^s}_{l^2(\Omega\times[0,T])}\right)\right\}+\lambda_1 \dfrac{\vertii{\Cb}_{l^2}^2}{2\sigma_1^2},\\\label{eqn:post_step2}
\text{s.t. }&\mcL_{K_{\Cb}} \text{ satisfies physics-based constraints.}
\end{align}
In the above formulation, the first two terms of the loss function are taken from the closed form of the negative log-posterior formulation in \eqref{eq:log-post}, and the last term provides regularization. This step is solved with the L-BFGS method. Then, based on the estimated solution $\Cb_2$, in step \textbf{1c)} we set ${\hat{\Cb}}:=\Cb_2$ consider the quadratic formulation which can be seen as an approximation of the first and last term in \eqref{eq:log-post}:
\begin{align}
\Cb_3:=\underset{\Cb}{\text{argmin}}&\sum_{s=1}^S\dfrac{\vertii{\tilde{u}_{NL,\Cb}^s-u_{DNS}^s}_{{l^2(\Omega\times[0,T])}}^2}{\vertii{u_{DNS}^s}_{{l^2(\Omega\times[0,T])}}^2}+\lambda \vertii{\Cb-\hat{\Cb}}_{l^2}^2,\label{eqn:1stepopt_new}\\
\text{s.t. }&\mcL_{K_{\Cb}} \text{ satisfies physics-based constraints,}
\end{align}
and can be solved as a linear problem. Here, the L-curve method is again employed with the loss function defined as $\mathcal{E}(\Cb{(\lambda)}) := \sum_{s=1}^S\frac{\vertii{\tilde{u}_{NL,\Cb{(\lambda)}}^s-u_{DNS}^s}_{l^2(\Omega\times[0,T])}^2}{\vertii{u_{DNS}^s}_{l^2(\Omega\times[0,T])}^2}$ and the regularization term as $ \mathcal{R}(\Cb{(\lambda)}):=\vertii{\Cb{(\lambda)}-\hat{\Cb}}_{l^2}^2$. With the estimated solution $\Cb_3$, we then update the regularization parameter for phase 2, which will be denoted as $\lambda_2$:
\begin{equation}\label{eqn:lambda_2}
\lambda_2:=\sum_{s=1}^S\dfrac{\vertii{u_{NL,\Cb_3}^s-u_{DNS}^s}_{l^2(\Omega\times[0,T])}^2}{\vertii{u_{NL,\Cb_3}^s}_{l^2(\Omega\times[0,T])}^2\vertii{\Cb_3-\Cb_2}_{l^2}^2}.
\end{equation}

Finally, with the initial values $(\Cb_2,\sigma_2)$ and the regularization parameter $\lambda_2$ from phase 1, we apply adaptive MCMC using the log-posterior \eqref{eq:log-post}, where the pair $(\hat{\Cb},\hat{\sigma})$ as well as the initial guess of $(\Cb,\sigma)$ are set as $(\Cb_2,\sigma_2)$. MCMC is implemented using the Uncertainty Quantification toolkit (UQTk)~\cite{DebusscherePCE:2004,DebusschereUQTk:2017}.
\section{Application to a Heterogeneous Elastic Bar}\label{sec:numerics}

In this section, we examine the efficacy of the proposed BNOR model and the two-phase learning algorithm, by considering the stress wave propagation problem described in Section ``Dispersion in Heterogeneous Materials''. Here, we seek a nonlocal homogenized model for the stress wave propagation in one-dimensional heterogeneous bars. Two exemplar heterogeneous bars are considered, one with periodic microstructure and one with random microstructure. For this problem, the goal is to obtain an effective nonlocal surrogate model from ground truth datasets generated by the DNS solver, acting at a much larger scale than the size of the microstructure. Since this problem has no ground-truth nonlocal kernel, we evaluate the surrogate by measuring its effectiveness in reproducing DNS data in applications that are subject to different loading conditions with a much longer time than the problems used as training data. To directly examine the extent to which our surrogate model reproduces the dispersion properties in the heterogeneous material, we also compare the group velocity curves from our model with the curves computed with DNS. Finally, we require that the learnt surrogate should provide a physically stable material model. To check this, we report the dispersion curve, whose positivity indicates that the learnt nonlocal model is physically stable. Besides the predicted MAP solutions, the Bayesian model also provides estimation of the predictive uncertainty for all above quantities of interests.

\subsection{Example 1: A Bar with Periodic Microstructure}\label{sec:periodic}

\paragraph{Data generation and settings.} First, we consider a heterogeneous bar with a periodic microstructure. As illustrated in the left top plot in Figure \ref{fig:bar}, the bar is a layered medium composed of two components, with the size of each layer $L=0.2$. Components 1 and 2 have the same density $\rho=1$ and Young's moduli $E_1=1$ and $E_2=0.25$, respectively. For the purpose of training and validation, we generate the DNS dataset with three types of data, and use the first two for training and the last one for validation of our algorithm. For all data we set the discretization parameters for the DNS solver as $\Delta t_{DNS}=0.01$, $\max\{\Delta x_{DNS}\}=0.01$, and consider the symmetric domain $\Omega = [-b, b]$. In what follows, $u$ represents the displacement, $v$ the velocity, and $f$ an external loading. The three types of data are chosen to follow a similar setting as in \cite{You2021}:

\medskip\noindent
\textit{Type 1: Oscillating source (20 observations).} We set $b=50$. The bar starts from rest such that $v(x,0)=u(x,0)=0$, and an oscillating loading is applied with $f(x,t) \!= e^{-\left(\frac{2x}{5kL}\right)^2} \!e^{-\left(\frac{t-t_0}{t_p}\right)^2}\!\cos^2\left(\frac{2\pi x}{kL}\right)$ with $k=1,2,\ldots,20$. Here we take $t_0=t_p=0.8$.

\medskip\noindent
\textit{Type 2: Plane wave with ramp  (11 observations).} We also set the domain parameter as $b=50$. The bar starts from rest ($u(x,0)=0$) and is subject to zero loading ($f(x,t)=0$). For the velocity on the left end of the bar, we prescribe  
\[v(-b,t)=
\left\{
\begin{aligned}
&\sin(\omega t)\sin^2\left(\frac{\pi t}{30}\right),&t\leq 15\\
&\sin(\omega t),&t>15
\end{aligned}
\right.
\] 
for $\omega=0.35,0.7,\cdots,3.85$.

\medskip\noindent
\textit{Type 3: Wave packet  (3 observations).} We consider a longer bar with $b=133.3$, with the bar starting from rest ($u(x,0)=0$), and is subject to zero loading ($f(x,t)=0$). The velocity on the left end of the bar is prescribed as
$v(-b,t)=\sin(\omega t)\exp{(-(t/5-3)^2)}$ with $\omega=2$, $3.9$, and $5$.

For all data types, parameters for the nonlocal solver and the optimization algorithm are set to $\Delta x= 0.05$, $\Delta t = 0.02$, $\delta =1.2$, and $M = 24$. For training purposes, we generate data of types 1 and 2 till $T = 2$. Then, to investigate the performance of our surrogate model in long-term prediction tasks we simulate till $T=100$ for data type 3.

\begin{figure}[h!]
\centering
\subfigure[First parameter]{\includegraphics[width=.30\columnwidth]{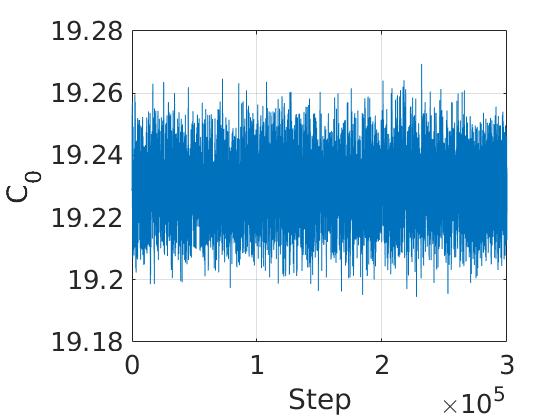}}
\subfigure[Second parameter]{\includegraphics[width=.30\columnwidth]{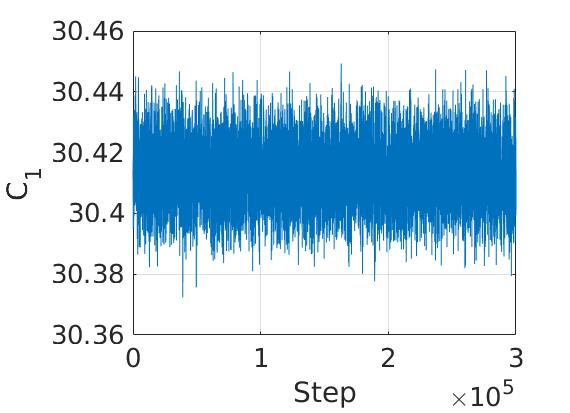}}
\subfigure[$\log(\sigma)$]{\includegraphics[width=.30\columnwidth]{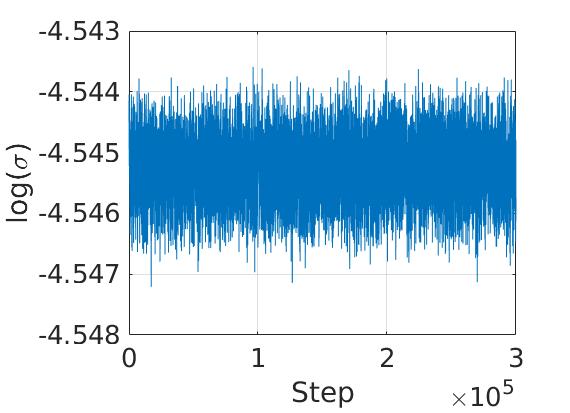}}
\caption{The first two kernel parameters $C_0$,  $C_1$ and $\log(\sigma)$ for a bar with periodic microstructure, plotted against the iteration number after eliminating the burn-in stage (the trace plot of MCMC).} 
\label{fig:periodic_traceplot}
\end{figure}

\begin{figure}[h!]
\centering
\subfigure[Optimal kernel]{\includegraphics[width=.45\columnwidth]{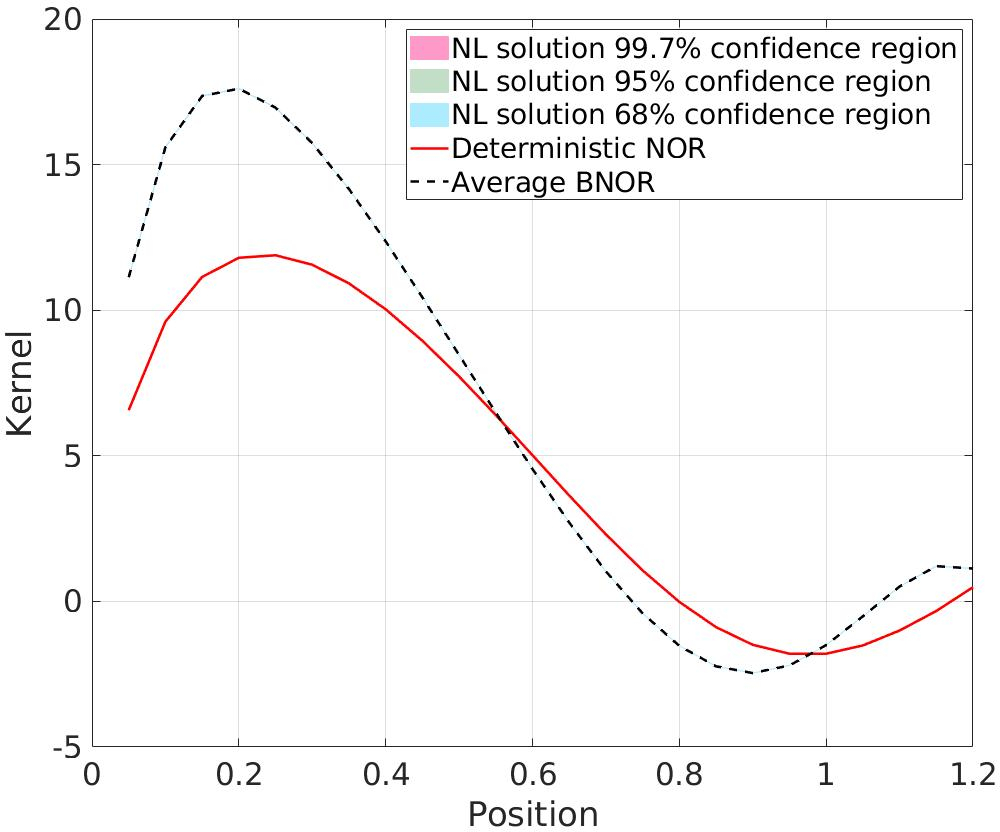}}\quad
\subfigure[Dispersion curve]{\includegraphics[width=.45\columnwidth]{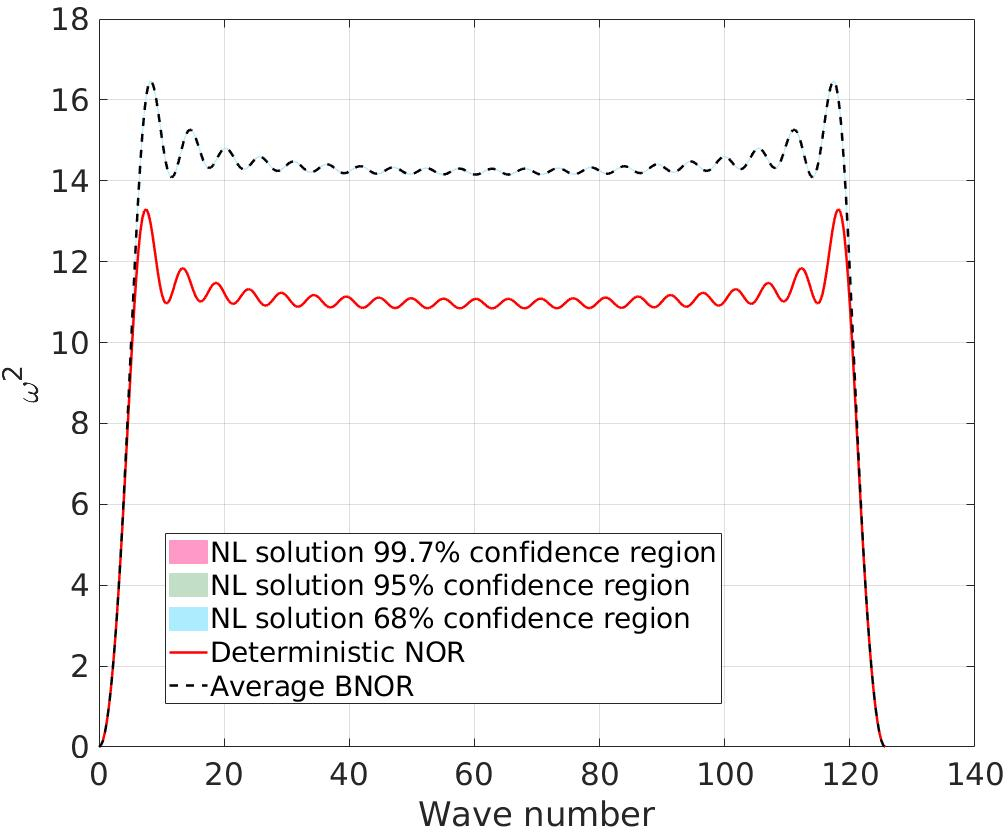}}
\subfigure[Group velocity]{\includegraphics[width=.45\columnwidth]{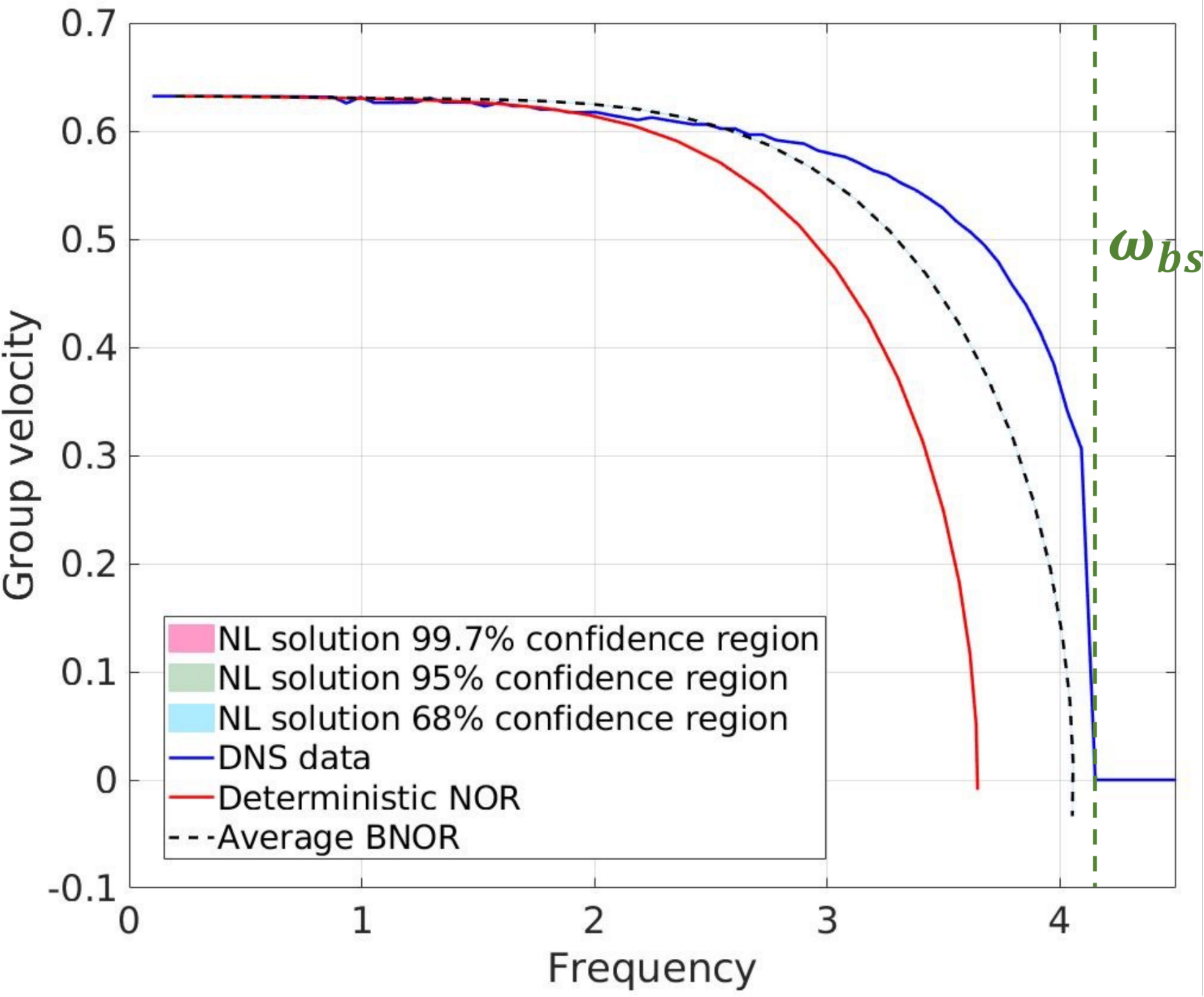}}
\caption{Optimal kernel, group velocity and dispersion curve for a bar with periodic microstructure. 
Here, the confidence region almost coincides with the curve of average BNOR, since the uncertainty in the kernel parameters is low.}
\label{fig:post_periodic}
\end{figure}

\paragraph{Results from Bayesian inference} An MCMC chain with 300,000 steps is generated, with approximately 30\% acceptance rate. To present the results, we post-process the chain based on 3,856 equally spaced samples, where the effective sample size (ESS) was calculated following the formulation in \cite{vats2019multivariate}. Figure \ref{fig:periodic_traceplot} shows a trace plot of the MCMC chain, illustrating good mixing of the chain. 
In Figure \ref{fig:post_periodic}(a), the nonlocal kernels using the learnt parameters are demonstrated, with predictive uncertainty ranges of $68\%$, $95\%$ and $99.7\%$ provided. 
Here, the confidence ranges were calculated based on the push-forward of the marginal posterior on $\Cb$ (using the samples from MCMC), through the nonlocal model \eqref{eqn:homo}. Both the averaged nonlocal kernel from the proposed BNOR algorithm and the deterministic nonlocal kernel from the original NOR are reported for comparison. 
Specifically, denoting the posterior mean of $\Cb$, as $\bar{\Cb}$, here the averaged nonlocal kernel is evaluated as $K_{\bar{\Cb}}\left(\frac{|\xi|}{\delta}\right)$. 
One can see that the averaged kernel from BNOR is close to the original NOR kernel when $\xi=|y-x|$ is relatively large ($>0.6$) with an almost negligible predictive uncertainty, while the discrepancy becomes more significant when $\xi$ approaches zero.  
However, the kernel values at $\xi\approx 0$ have a relatively small impact on the nonlocal operator $\mcL_{K_{\Cb}}$, and consequently also on the corresponding nonlocal solution $u_{NL}$ as well as the loss function in our optimization problems. 
Moreover, sign-changing behaviors are observed on both kernels, which are consistent with literature \cite{xu2020deriving,you2020data}. Then, in Figure \ref{fig:post_periodic}(b) we illustrate the deterministic and mean estimates, and predictive uncertainty ranges, of the dispersion curve. Here, we can see that both the averaged model from BNOR and the deterministic model from NOR possess positivity, indicating that both models correspond to a family of physically stable material models. Finally, in Figure \ref{fig:post_periodic}(c) we plot the estimated group velocity against frequency $\omega$, in comparison with the group velocity profile from the DNS simulation. These results indicate that both the averaged model from BNOR and the deterministic model from original NOR are able to match the DNS behavior for low frequency values $\omega$. 
When the frequency is getting closer to the band stop, $\omega_{bs}$, comparing with the deterministic model from original NOR, we notice that the averaged model from BNOR has reproduced the group velocity of the DNS model with better accuracy. Here, we stress that since the method for evaluating the DNS $v_g$ from wave packets is susceptible to error near the band stop, it is generally infeasible to expect our nonlocal surrogate model to match perfectly with the DNS group velocity. For both the dispersion curve and group velocity, predictive uncertainties are again negligible.

\begin{figure}[htp]
    \centering
    \includegraphics[width=1.0\columnwidth]{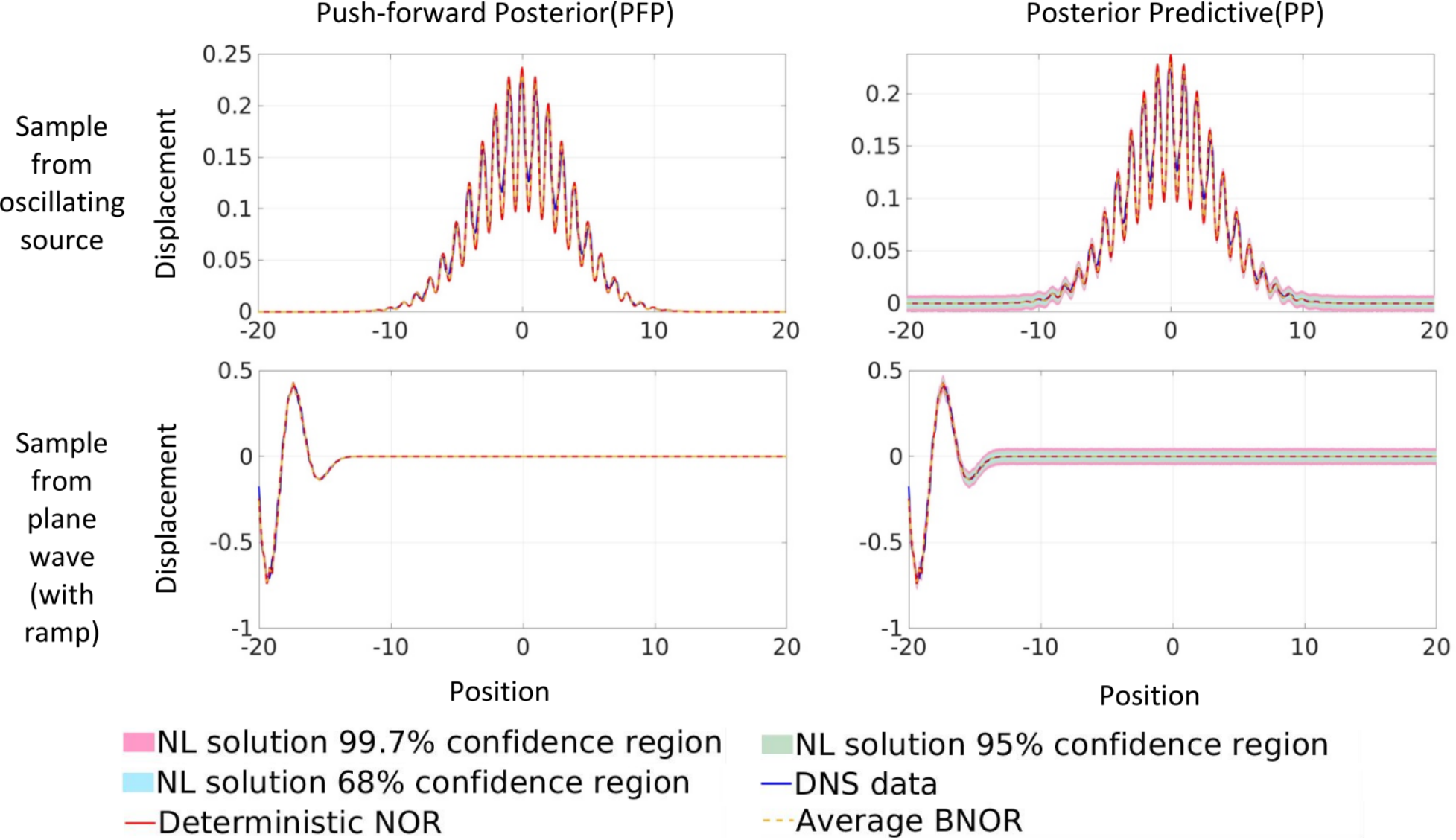}
    \caption{Comparison of two posterior uncertainty results for the prediction of two exemplar examples from the training dataset, on the periodic microstructure case. Left: displacement field prediction and confidence regions from the push-forward posterior (PFP) approach. Right: displacement field prediction and confidence regions from the posterior predictive (PP) approach.}
    \label{fig:periodic_tr}
\end{figure}

We note in particular that the evident negligible level of predictive uncertainty in these results is due to the fact that (1) our error model is rather simple, presuming independent noise, such that each additional data point adds information, and (2) we have a large number of data points. The combination of both factors results in the posterior on $\Cb$ being highly concentrated, thus exhibiting minimal uncertainty, even though the discrepancy between the two models is non-negligible. In order to endow the nonlocal predictive model with uncertainties that better approximate the discrepancy between it and the high-fidelity model, one can resort to a more elaborate statistical construction to represent model error~\cite{Kennedy:2000}, particularly as embedded in the model construction~\cite{Sargsyan:2015,Sargsyan:2019}. We reserve this pathway forward for future work.

We illustrate further, in Fig.~\ref{fig:periodic_tr}, posterior uncertainty on two predicted solutions that are part of the training data. In each case, we highlight two posterior uncertainty results. The first, commonly termed the push-forward posterior (PFP), is the push-forward of the marginal posterior $p(\Cb|D)$ through the non-local model $u_{NL,\Cb}(x,t)$. The second, termed the posterior predictive (PP), is the push-forward of the full posterior $p(\Cb,\sigma|D)$ through the data model $u_{NL,\Cb}(x,t)+\epsilon$. The PFP provides the posterior uncertainty on the model predictions given posterior knowledge of $\Cb$. On the other hand, the PP provides a posterior density on predictions of the data model, which serves as a diagnostic measure of the quality of this model, composed of the physical model and the proposed error model, as a predictor of the data\footnote{Note that, in an additive model error framework~\cite{Kennedy:2001} where the $\epsilon$ term, defined to include some correlation structure, would be included in ``corrected" model, such that the associated PP \& PFP are equivalent.}. It is expected that the PP would exhibit higher uncertainty on predictions than the PFP. Further, an ideal PP should span the data, and its samples ought to be statistically indistinguishable from the data. Considering the results in Fig.~\ref{fig:periodic_tr}, we can see that the BNOR solution with the mean $\Cb$, the deterministic NOR, and the DNS solution are all very well matched for the plane wave case, with discernible differences in the displacement magnitudes in the oscillating source case. Further, we see that the PFP results in negligible predictive uncertainty, as already observed in Fig.~\ref{fig:post_periodic} for the kernel and associated quantities of interest. On the other hand, the PP results exhibit non-negligible uncertainty, particularly at low mean-output levels for positions beyond $\pm10$ for the oscillating source case, and above position -15 for the plane wave case. However, we find that the PP uncertainty is again negligible at high signal levels around position 0 in the oscillating source case, and below position -15 in the plane wave case. Moreover, the PP uncertainty is small relative to the discrepancy between the model and the data at large signal levels in the oscillating source case, while it is larger than the discrepancy at low signal levels. Clearly, the PP does not do an ideal job of spanning the discrepancy from the data, indicating that a more flexible/accurate, perhaps embedded, error model would be needed to better capture the data error.

\paragraph{Validation on wave packet}

In order to further test the predictive performance of our surrogate model, we now demonstrate its capability in reproducing DNS simulations through the prediction error of $u$ on data type 3, involving wave packet problems where a long time simulation is conducted on a long bar, and the loading scenario is substantially different from the training cases. With the proposed BNOR approach, for each loading scenario both the prediction of displacement field $u_{NL,\Cb}(x_i,t^n)$ and the $68\%-95\%-99.7\%$  confidence regions are provided. 
Specifically, we present both the PFP $p(u_{NL,\Cb}(x_i,t^n)|D)$ and PP $p(u_{NL,\Cb}(x_i,t^n)+\epsilon|D)$ results, where $\epsilon\sim \mathcal{N}(0,\sigma^2||u_{NL,\Cb}(x,t)||^2_{l_2([-b,b]\times[0,T])})$. 

\begin{figure}[htp]
    \centering
    \includegraphics[width=1.0\columnwidth]{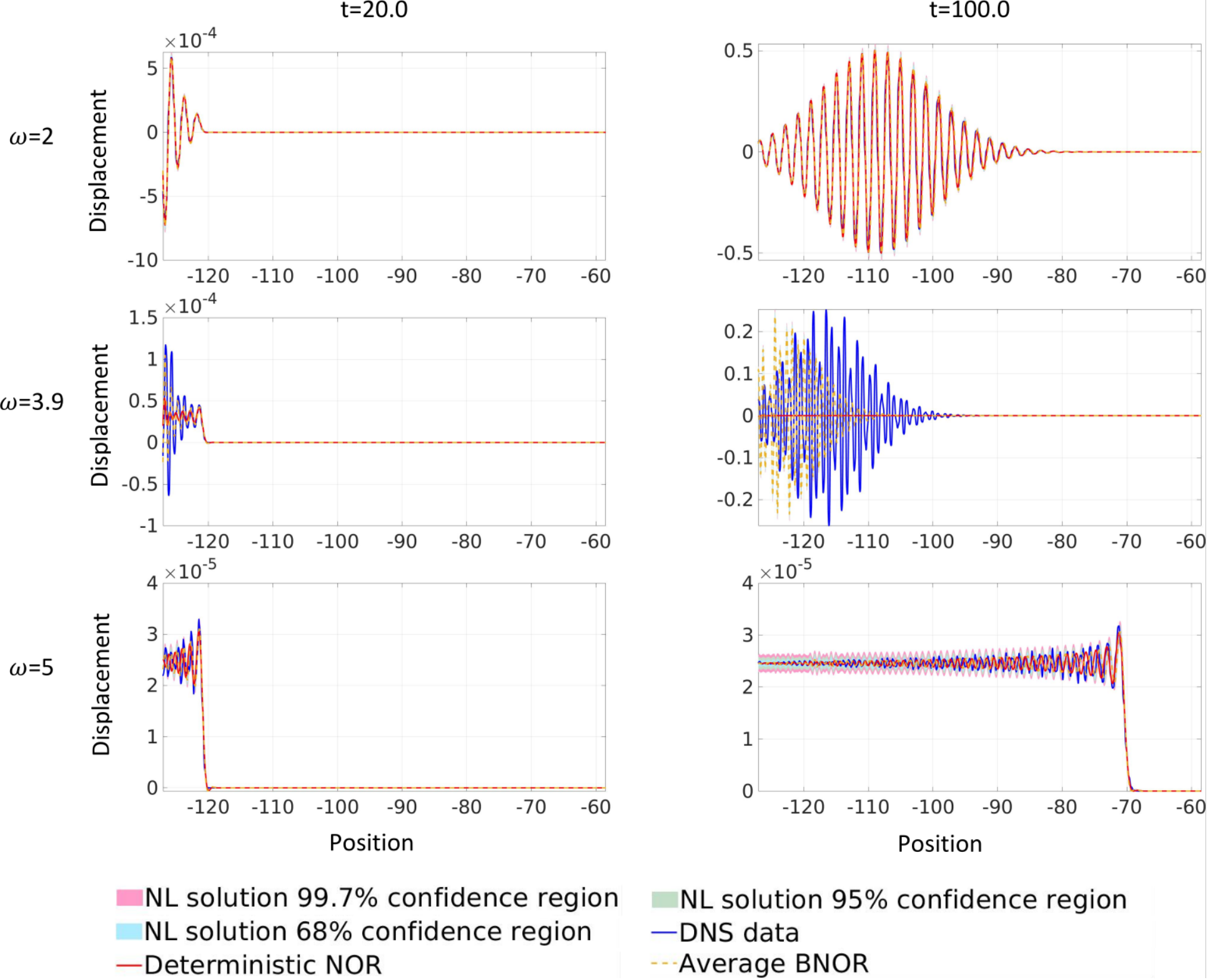}
    \caption{Validation on the wave packet traveling problem for a bar with periodic microstructure, illustrating PFP uncertainty. Plots from top to bottom are corresponding to different loading frequencies: (top) $\omega=2<\omega_{bs}$, (middle) $\omega=3.9\approx\omega_{bs}$, and (bottom) $\omega=5>\omega_{bs}$. Left column shows the simulation results on a relatively short time ($T=20$), and the right column demonstrates long time simulation results ($T=100$). 
    }
    \label{fig:periodic_wp_nosig}
\end{figure}

\begin{figure}[htp]
    \centering
    \includegraphics[width=1.0\columnwidth]{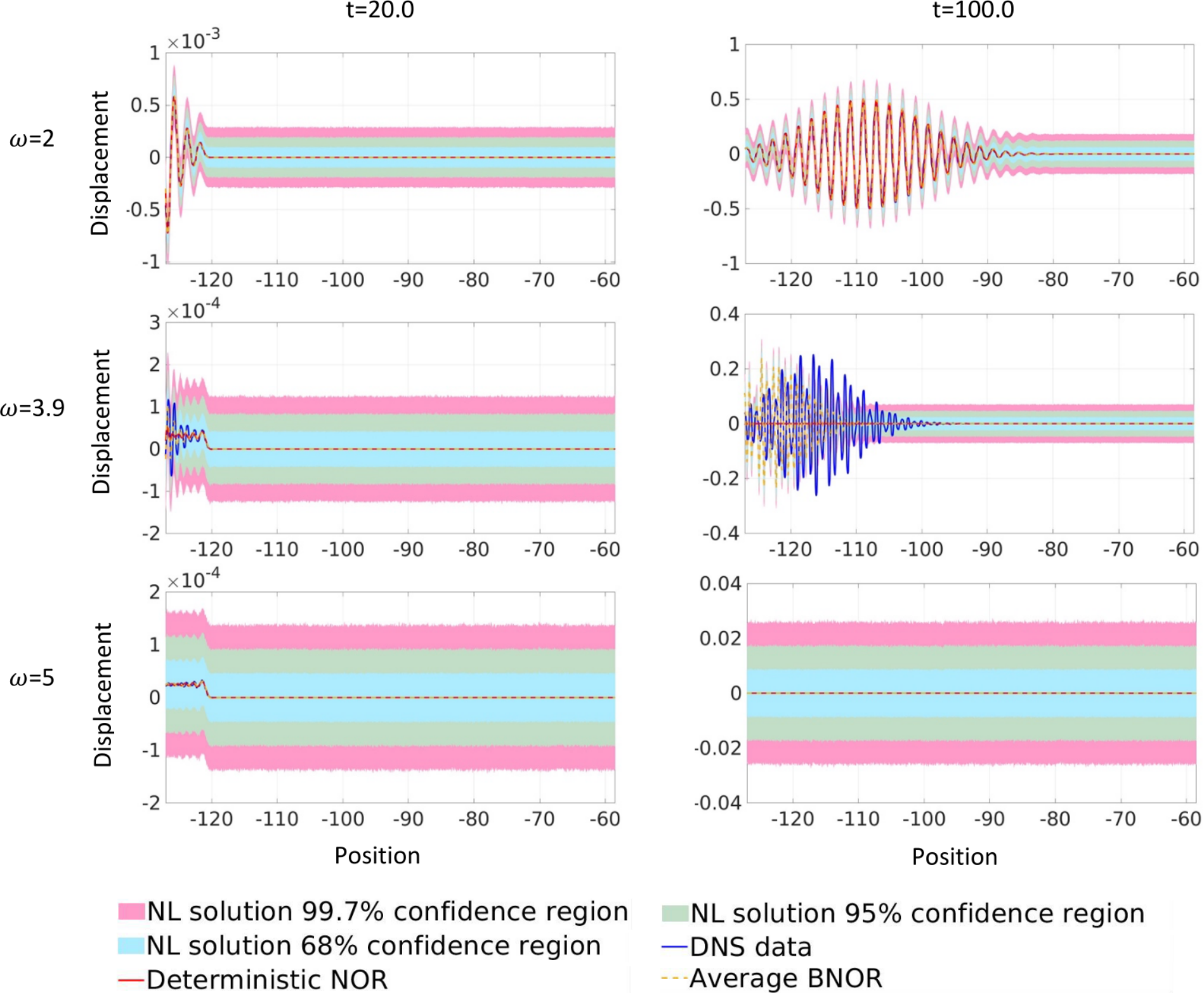}
    \caption{Validation on the wave packet traveling problem for a bar with periodic microstructure, illustrating PP uncertainty. Plots from top to bottom are corresponding to different loading frequencies: (top) $\omega=2<\omega_{bs}$, (middle) $\omega=3.9\approx\omega_{bs}$, and (bottom) $\omega=5>\omega_{bs}$. Left column shows the simulation results on a relatively short time ($T=20$), and the right column demonstrates long time simulation results ($T=100$). 
    }
    \label{fig:periodic_wp}
\end{figure}

In Figures \ref{fig:periodic_wp_nosig} and \ref{fig:periodic_wp} we consider solutions corresponding to three values of frequency $\omega$: $\omega_1=2<\omega_{bs}$, $\omega_2=3.9\approx\omega_{bs}$ and $\omega_3=5>\omega_{bs}$, respectively. Further, results in the two figures illustrate PFP (Fig.~\ref{fig:periodic_wp_nosig}) and PP (Fig.~\ref{fig:periodic_wp}) uncertainty as indicated. For the the first two values of $\omega$, the exact stress wave is anticipated to be traveling in time. For the last case, since the value of $\omega$ is beyond the band stop and corresponds to a zero DNS group velocity, the exact wave does not travel in time. Both short time ($t=20$) and long time ($t=100$) predictions are considered for each case. The deterministic solution from the standard NOR algorithm and the DNS data are also reported for comparison. We observe at the outset the negligible PFP uncertainty (Fig.~\ref{fig:periodic_wp_nosig}), as already observed in Fig.~\ref{fig:periodic_tr} for training data predictions, while larger PP uncertainty is evident (Fig.~\ref{fig:periodic_wp}), and quite prominently as the displacement goes to zero.

Further, for the loading frequency $\omega=2.0$ case, we observe good agreement between the mean of the nonlocal solutions and the DNS data for both $t=(20,100)$ cases.  Thus, the optimal nonlocal surrogate corresponding to our learnt samples can accurately reproduce both short and long-time stress wave propagation in this frequency range. Here, we stress that although we refer to $t=20$ as a relative short term, it is still much longer ($10\times$) than the total time interval our training algorithm has seen from the training samples. These findings illustrate the generalization property of our algorithm in extrapolation tasks in this frequency range.
On the other hand, considering the wave propagation results for the loading frequency $\omega=3.9$ case, which is close to the band stop frequency, a larger discrepancy between the DNS solution and the predicted displacement field is observed, especially in the region near the wave front. More specifically, at $t=20$, the mean of the nonlocal solutions is slightly off from the DNS solution, while the latter still lies inside the $68\%$ PP confidence region (Fig.~\ref{fig:periodic_wp}). This discrepancy grows appreciably as time moves forward, as can be seen for the $t=100$ case, extending well beyond the PP range. Clearly, our extrapolative behavior is poor for loading frequencies in the vicinity of the band stop frequency.
Finally, when the loading frequency $\omega=5.0>\omega_{bs}$ and the wave barely propagates, as shown in both figures, our model successfully captures the phenomenon that the wave stops traveling.

Generally, based on these results, we can say that the proposed nonlocal surrogate model performs well in short term extrapolative prediction tasks, and the posterior predictive associated with the current additive error model provides an adequate coverage of the DNS data, even in extrapolation, when the loading frequency is away from the band stop frequency, $\omega_{bs}$. To further improve the prediction accuracy and the model discrepancy representation, more training observations on a longer time and/or a more flexible and sophisticated error model employing a Gaussian process construction should help, allowing always for the lack of expected accuracy from the statistical model in extrapolation if the mean model itself is failing.

\subsection{Example 2: A Bar with Random  Microstructure}\label{sec:disorder}

\paragraph{Data generation and settings.}
We now consider a heterogeneous bar with random microstructure, where the size of each layer is now defined by a random variable. In particular, the density for both components is still set as $\rho=1$ and the Young's moduli are $E_1=1$ and $E_2=0.25$. The layer sizes $L_1$, $L_2$ are two random variables, both satisfying a uniform distribution: $L_1,L_2\sim \mathcal{U}[(1-D){L},(1+D){L}]$, with the averaged layer size ${L}=0.2$ and the disorder parameter $D=0.5$. To generate the training dataset, we simulate the wave propagation using the DNS solver, under the same settings as in data type 1 and data type 2 of the periodic bar case. Then, the wave packet problem is again considered for the purpose of validation. Comparing with the periodic microstructure case, from the group velocity generated by the DNS simulations (see Figure \ref{fig-vgtypical} bottom plot) we note that the band stop generally occurs at a lower frequency in the random microstructure case. In fact, for the microstructure considered in this bar, an estimated band stop frequency $\omega_{bs}\approx 3$ can be obtained from the DNS simulations. Thus, in this section, wave packets with frequencies $\omega=1,2,3$ and $4$ are considered as the validation samples, with the purpose of investigating the performance of our nonlocal surrogate model when the loading frequencies are below ($\omega=1,2$), around ($\omega=3$), and above ($\omega=4$) the estimated band stop frequency $\omega_{bs}$. 

\begin{figure}[h!]
\centering
\subfigure[First parameter]{\includegraphics[width=.30\columnwidth]{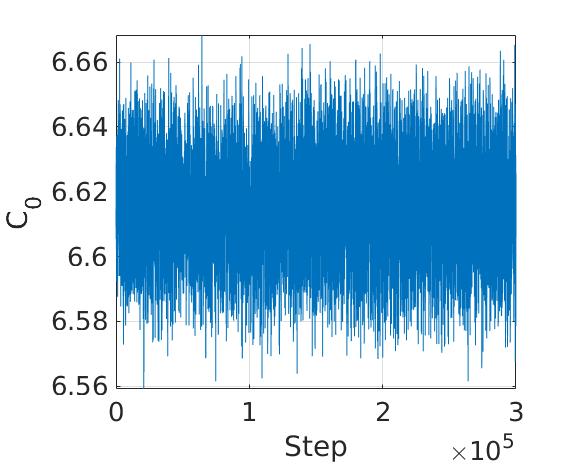}}
\subfigure[Second parameter]{\includegraphics[width=.30\columnwidth]{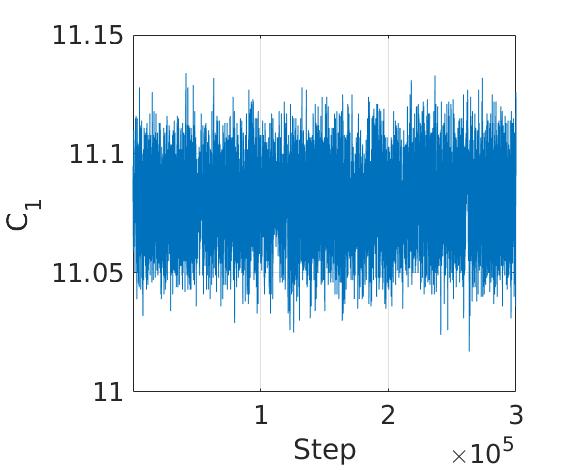}}
\subfigure[$\log(\sigma)$]{\includegraphics[width=.30\columnwidth]{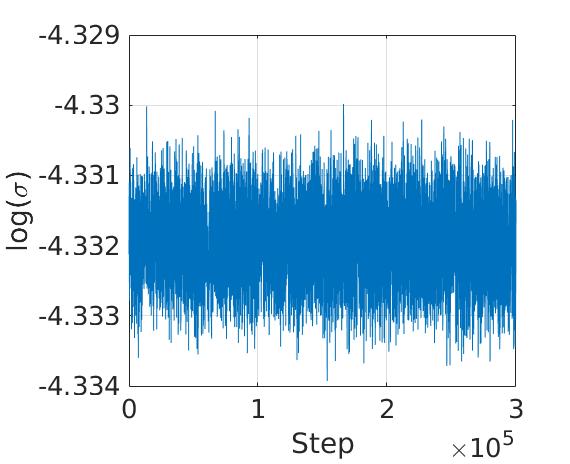}}
\caption{The first two kernel parameters $C_0$,  $C_1$ and $\log(\sigma)$ for a bar with random microstructure, plotted against the iteration number after eliminating the burn-in stage (the trace plot of MCMC).}
\label{fig:disorder_traceplot}
\end{figure}

\paragraph{Results from Bayesian inference.} We run an MCMC chain with 3$\times 10^5$ steps, with a $31\%$ acceptance rate, and post-process it using 4,212 equally spaced samples guided by an estimated ESS. 
The ESS, the trace plot, the predicted averaged kernel, and its dispersion properties are all calculated similarly as for the periodic bar. Figure \ref{fig:disorder_traceplot} shows the exemplar trace plot of the MCMC chain for two kernel parameters, $C_0$ and $C_1$, and $\sigma$, illustrating good mixing. Compared with the periodic bar, slightly larger values of $\sigma$ are obtained here. Hence, the randomness in material microstructure is anticipated to induce a larger model discrepancy.
%
%
\begin{figure}[h!]
\centering
\subfigure[Optimal kernel]{\includegraphics[width=.45\columnwidth]{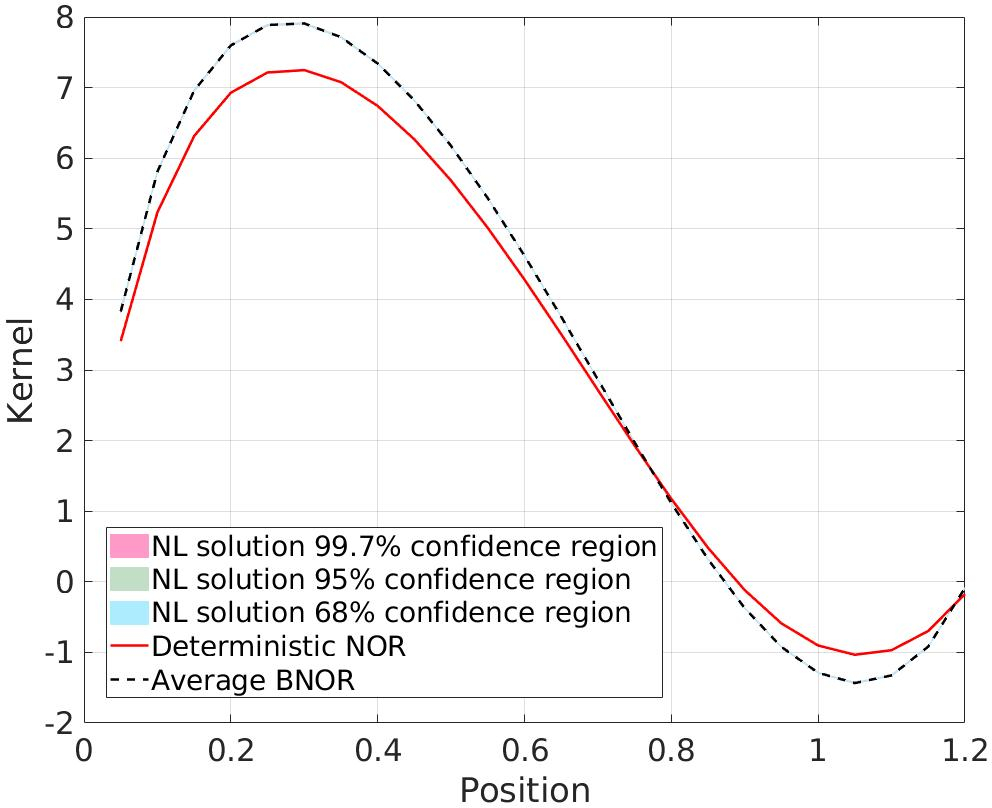}}\quad
\subfigure[Dispersion curve]{\includegraphics[width=.45\columnwidth]{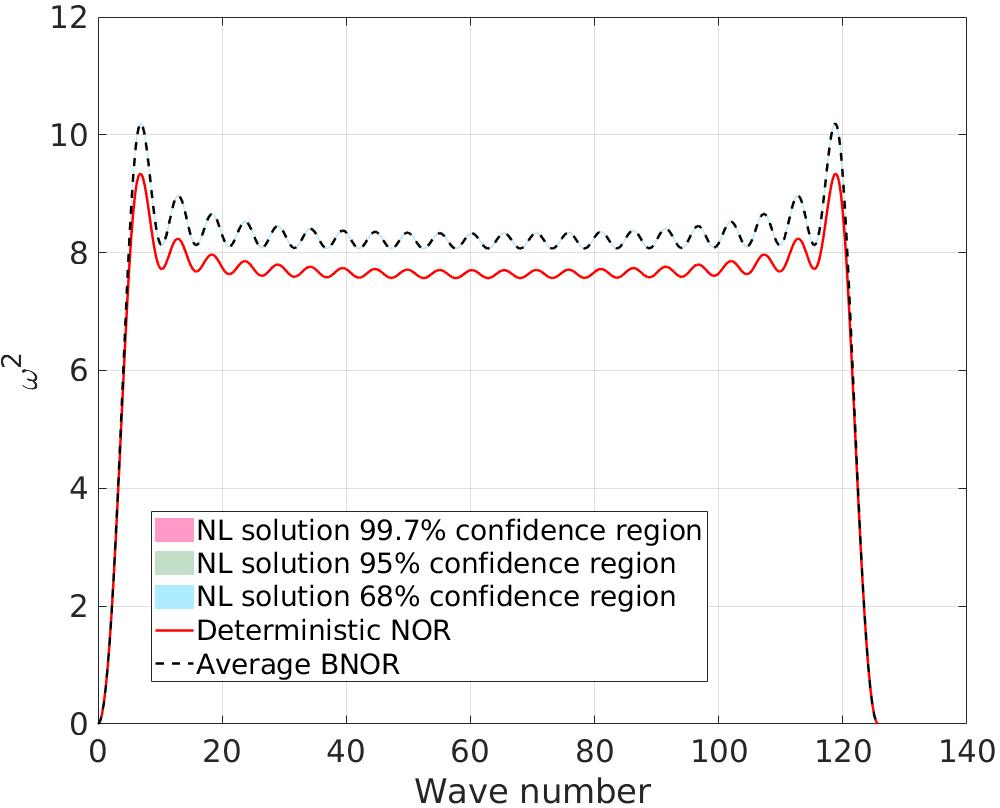}}
\subfigure[Group velocity]{\includegraphics[width=.45\columnwidth]{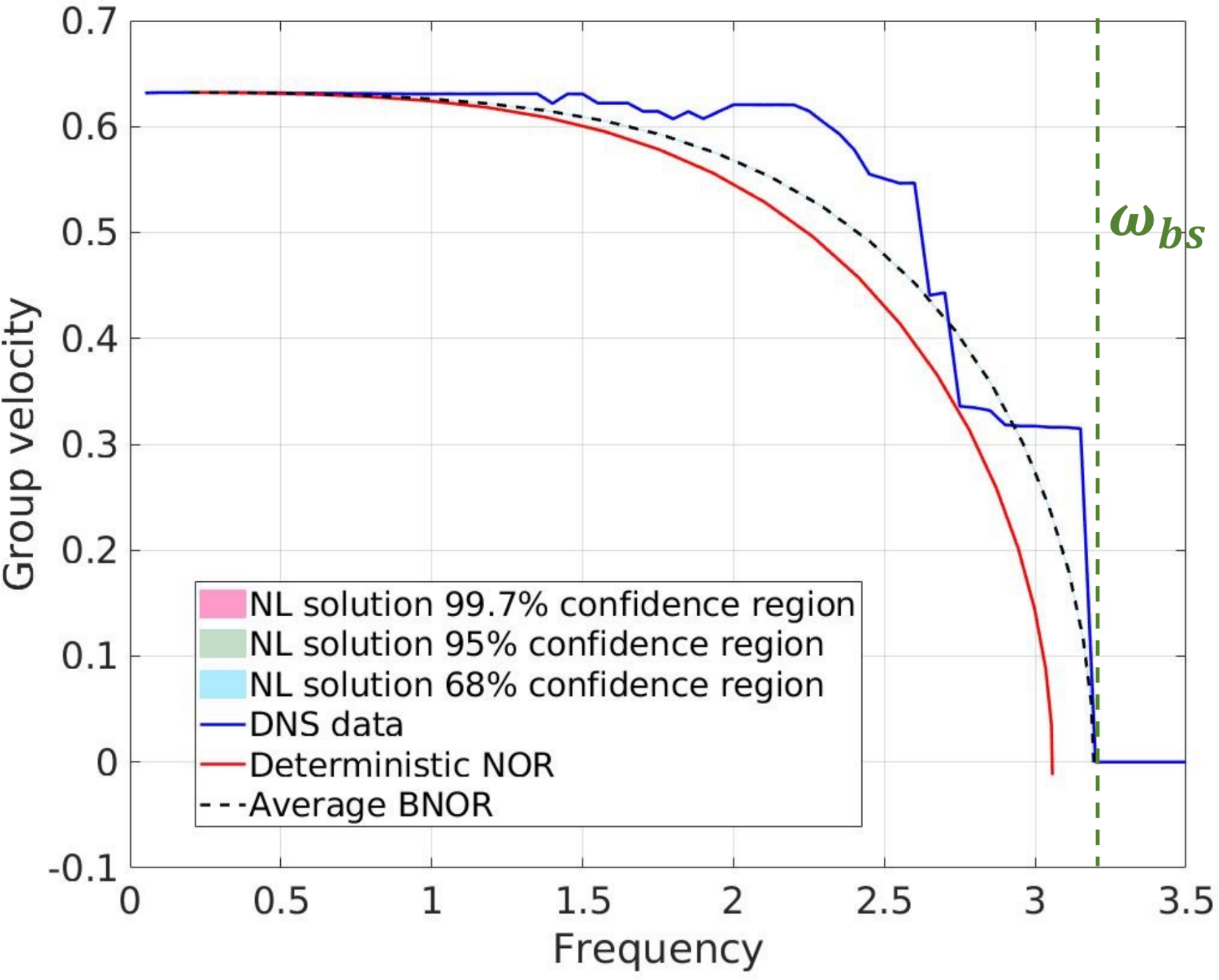}}
\caption{Optimal kernel, group velocity and dispersion curve for a bar with random microstructure. 
Since the uncertainty in the kernel parameters remains low, the confidence regions generally coincide with the curve of average BNOR.}
\label{fig:post_disorder}
\end{figure}
Then, in Figure \ref{fig:post_disorder} we plot the predicted averaged nonlocal kernel, the dispersion curve, and the group velocity profiles, together with the $68\%-95\%-99.7\%$ PFP confidence regions. The results from original NOR algorithm are also reported for comparison. For the estimated nonlocal kernels, as demonstrated in Figure \ref{fig:post_disorder}(a) we observe a small discrepancy between the averaged kernel and the deterministic kernel from NOR, together with very small confidence regions. These trends are also observed in Figure \ref{fig:post_disorder}(b) and Figure \ref{fig:post_disorder}(c), for the dispersion curves and group velocity profiles, respectively. As in the periodic bar case, for this random microstructure the predicted nonlocal model from BNOR again possesses physical stability, and successfully identifies the band stop. In Figure \ref{fig:post_disorder}(c), a relatively larger discrepancy is observed between the group velocity from the DNS solver and profile from the estimated nonlocal surrogate, possibly due to the fact that the material randomness introduces larger errors in the approximated DNS group velocity, $v_g$, especially for the frequencies near the band stop, as discussed in Section ``Dispersion in Heterogeneous Materials''. 



\begin{figure}[htp]
    \centering
    \includegraphics[width=1.0\columnwidth]{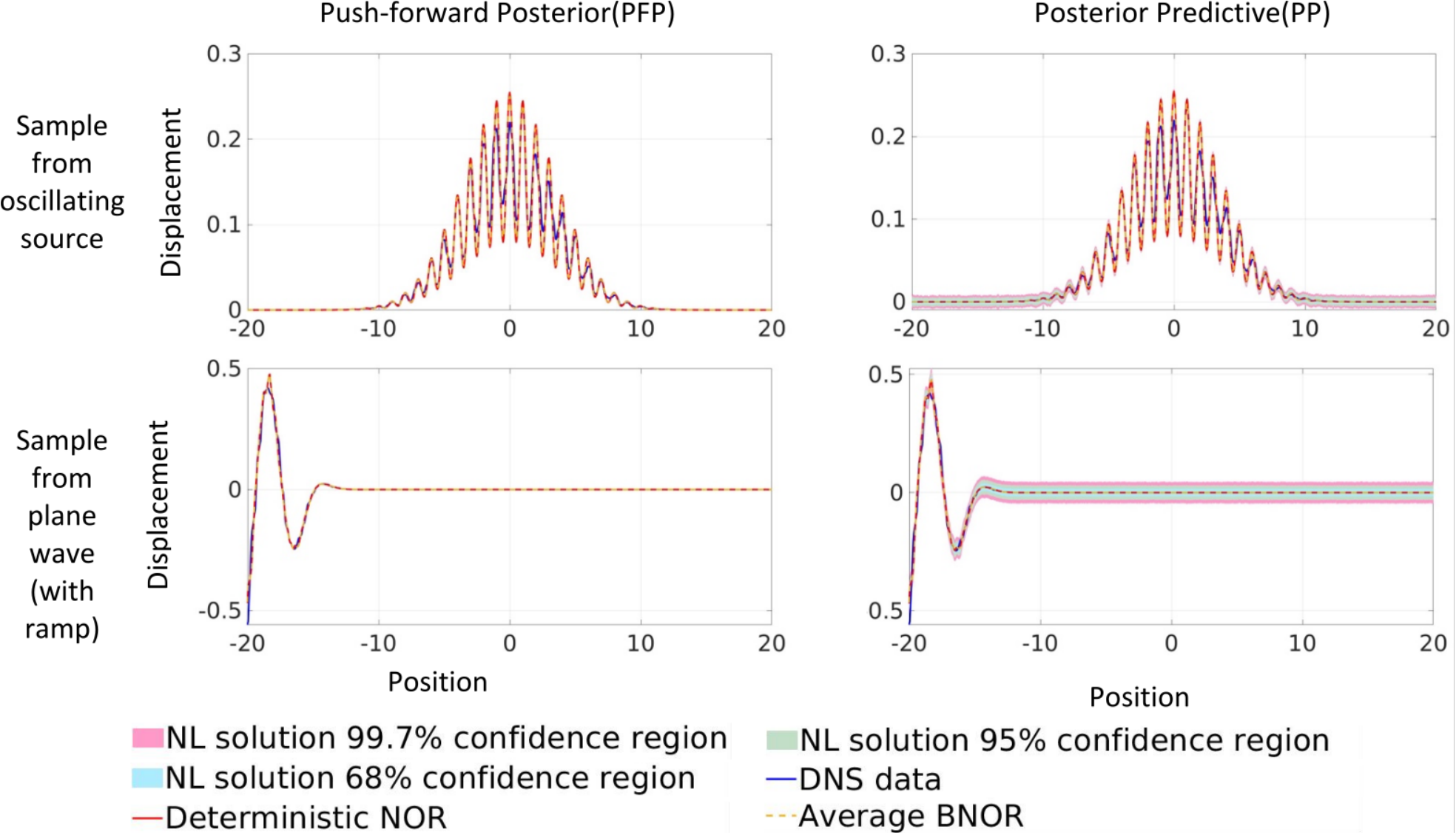}
    \caption{Comparison of two posterior uncertainty results for the prediction of two exemplar examples from the training dataset, on the random microstructure case. Left: displacement field prediction and confidence regions from the push-forward posterior (PFP) approach. Right: displacement field prediction and confidence regions from the posterior predictive (PP) approach.}
    \label{fig:disorder_tr}
\end{figure}

We illustrate again, in Fig.~\ref{fig:disorder_tr}, posterior uncertainty on two predicted solutions that are part of the random microstructure training data, including both the PFP and PP posterior uncertainty results. 
We can see that the BNOR solution with the mean $\Cb$, the deterministic NOR, and the DNS solution are all very well matched for the plane wave case. On the other hand, again, the oscillating source case exhibits discernible displacement magnitude differences between the two nonlocal solutions and the DNS solution. We also see that the PFP results in negligible predictive uncertainty, as already observed in Fig.~\ref{fig:post_disorder}. On the other hand, the PP results exhibit non-negligible uncertainty at low mean-output levels for both the oscillating source and plane wave case, while exhibiting again negligible uncertainty at high displacement levels even where the discrepancy with the DNS data is large. Here again, the PP with the current data model does not span the discrepancy from the data well, and is a candidate for improvement in future work.

\begin{figure}[htp]
    \centering
    \includegraphics[width=1.0\columnwidth]{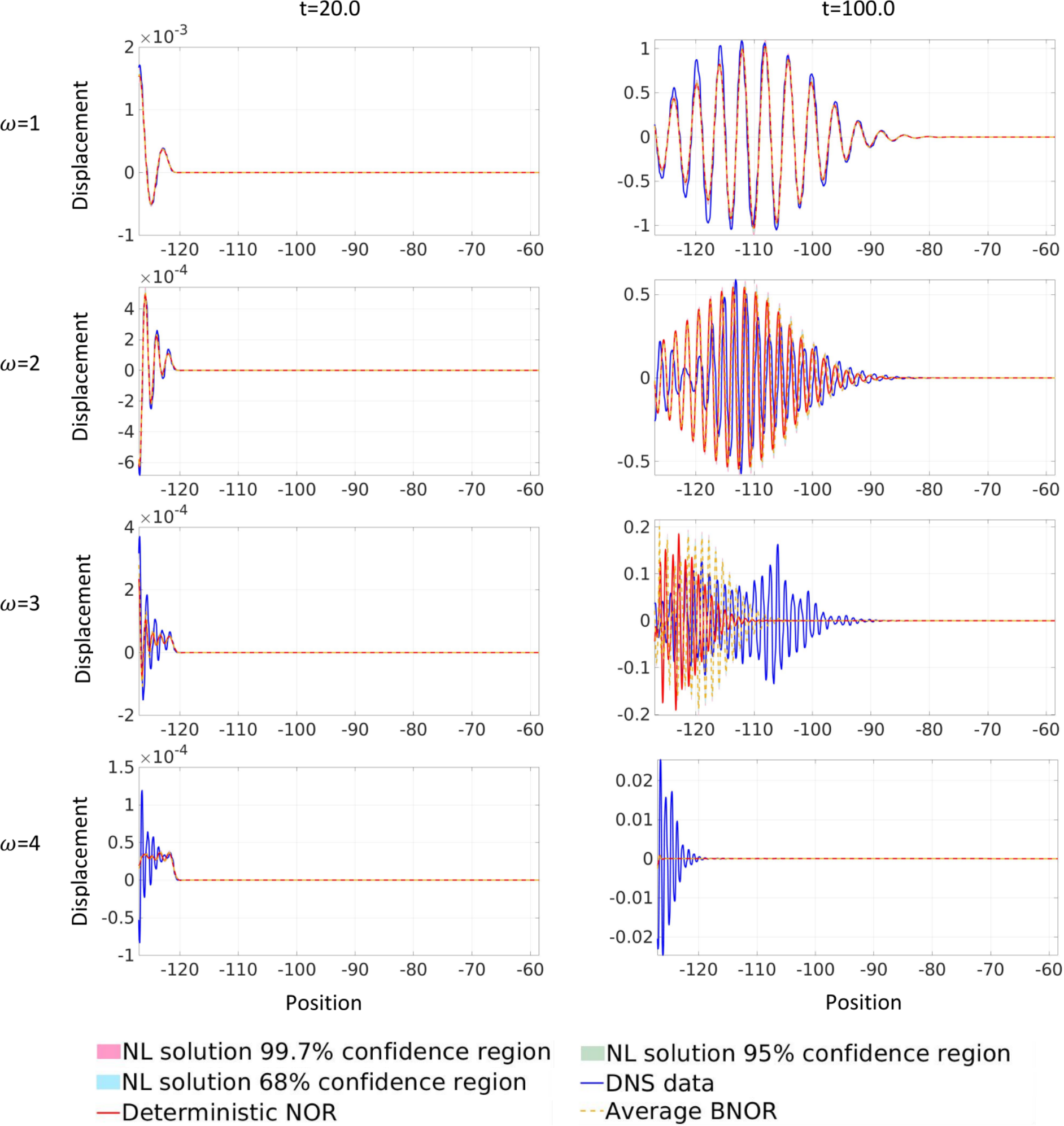}
    \caption{Validation on the wave packet traveling problem for a bar with random microstructure, illustrating PFP uncertainty. Plots from top to bottom are corresponding to different loading frequencies: 1) $\omega=1<\omega_{bs}$, 2) $\omega=2<\omega_{bs}$, 3) $\omega=3\approx\omega_{bs}$, and 4) $\omega=4>\omega_{bs}$. Left column shows the simulation results on a relatively short time ($T=20$), and the right column demonstrates long time simulation results ($T=100$).}
    \label{fig:disorder_wp_nosig}
\end{figure}

\begin{figure}[htp]
    \centering
    \includegraphics[width=1.0\columnwidth]{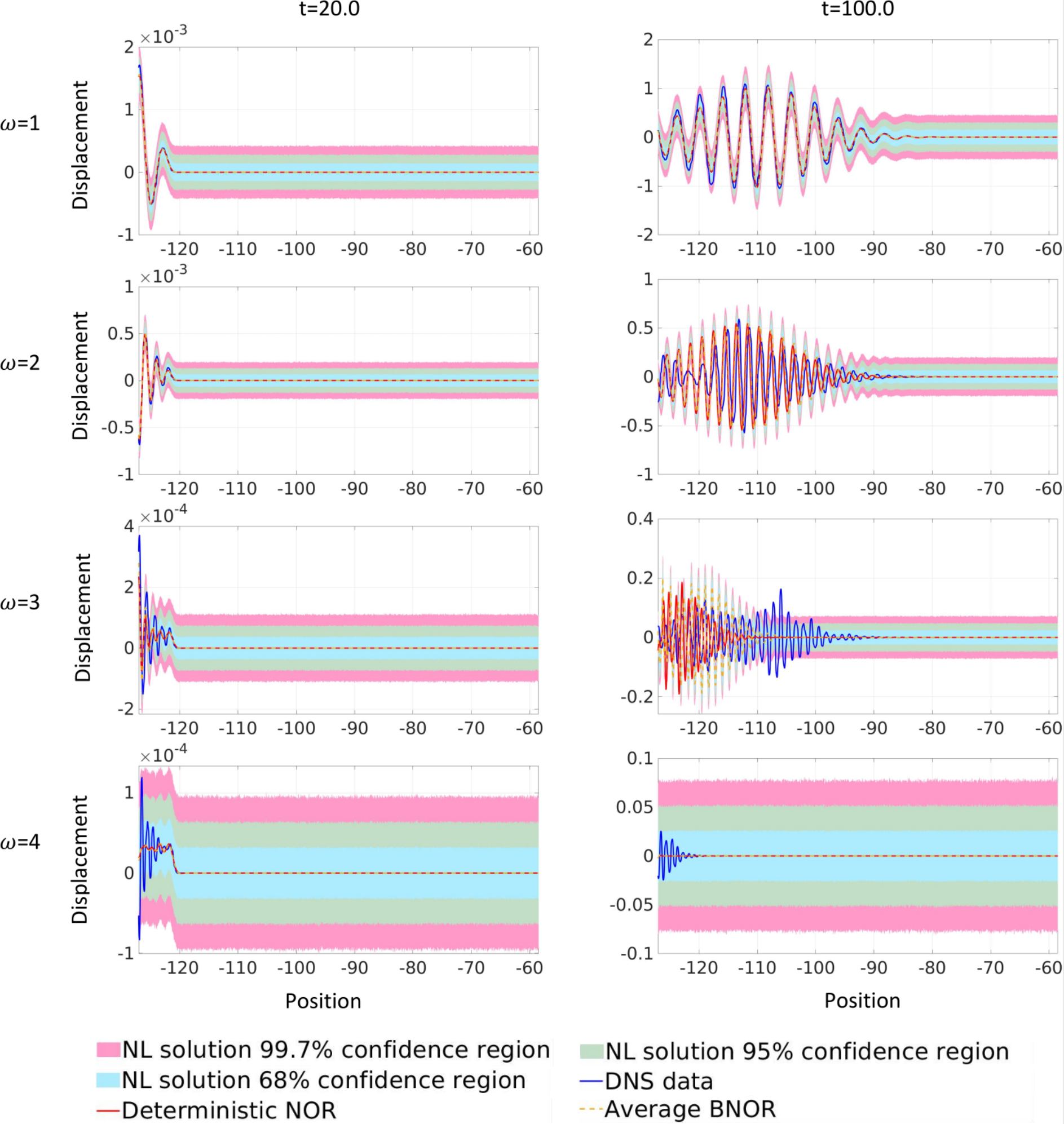}
    \caption{Validation on the wave packet traveling problem for a bar with random microstructure, illustrating PP uncertainty. Plots from top to bottom are corresponding to different loading frequencies: 1) $\omega=1<\omega_{bs}$, 2) $\omega=2<\omega_{bs}$, 3) $\omega=3\approx\omega_{bs}$, and 4) $\omega=4>\omega_{bs}$. Left column shows the simulation results on a relatively short time ($T=20$), and the right column demonstrates long time simulation results ($T=100$).}
    \label{fig:disorder_wp}
\end{figure}

\paragraph{Validation on wave packet}
Following the same procedure as for the periodic microstructure case, here we again demonstrate the generalization capability of the learnt nonlocal surrogate model on different domains, boundary conditions, and longer simulation time, by considering the wave packet problem. Prediction results subject to four loading frequencies, $\omega=1.0<\omega_{bs}$, $\omega=2.0<\omega_{bs}$, $\omega=3.0\approx\omega_{bs}$, and $\omega=4.0>\omega_{bs}$, are provided on a relatively short time ($t=20$) and a longer time ($t=100$) simulations. The validation results are illustrated in Figures \ref{fig:disorder_wp_nosig} and \ref{fig:disorder_wp} illustrating PFP and PP uncertainty respectively. The deterministic nonlocal solution from the original NOR algorithm and the DNS data are also reported for comparison. 

We can see here again the negligible PFP uncertainty, even under extrapolation, in Fig.~\ref{fig:disorder_wp_nosig}, while significant PP uncertainty is evident in Fig.~\ref{fig:disorder_wp}.
Further, for the short time prediction task ($t=20$), a good agreement between the mean of the uncertain nonlocal solution and the DNS data is observed for $\omega=1,2$, validating the generalizability of the learnt nonlocal surrogate under these loading conditions up to this time. However, there is clear mismatch with DNS for $\omega=3,4$ in the region near the wave front. 
For long time prediction ($t=100$), the extrapolated learnt nonlocal surrogate model exhibits discernible differences from the DNS solution under all loading cases, with the minimal differences being evident at $\omega=1$. At this loading the differences are confined to small magnitude differences in the oscillatory trace. On the other hand, more significant differences in both magnitude and phase are evident for $\omega=2,3,4$. 
Qualitatively, a satisfactory result is indeed obtained when the loading frequency $\omega=4.0>\omega_{bs}$ in that the wave barely travels in time. 
Except for the case with $t=100$ and $\omega=3$, the DNS solution always lies inside the $68\%$ PP confidence region, highlighting the efficacy of our additive error model under these conditions in extrapolation. 

\section{Conclusion}\label{sec:conclusion}

In this paper, we have proposed BNOR -- an approach for learning the optimal nonlocal surrogate together with modeling discrepancy characterization for heterogeneous material homogenization. Our work is built based on the nonlocal operator regression approach \cite{you2020data}, which aims to provide a well-posed and generalizable nonlocal surrogate model from high-fidelity simulations and/or experimental measurements of the displacement fields, and allows for accurate simulations at a larger scale than the microstructure. Because of these desired properties, the nonlocal surrogate model from NOR is readily applicable for unseen prediction tasks, such as to find the solutions at much larger times than the time instants used for training, and on problem settings that are substantially different from the training data set. On the other hand, since the nonlocal surrogate model from NOR serves as a homogenized surrogate of the original complex physical system without resolving its heterogeneities at the microscale, unavoidable modeling discrepancy will be introduced, which contribute to the overall prediction error and uncertainty.

To quantify the model predictions and associated uncertainty in future prediction tasks, in this work we used an independent additive Gaussian data model centered on the NOR approach, in order to represent and quantify the homogenization modeling discrepancies. The framework is developed within a Bayesian inference context, where NOR model parameters are inferred simultaneously along with parameters that characterize errors relative to the data. To solve the Bayesian inference problem efficiently, a two-phase MCMC algorithm is proposed, enabling an efficient and non-intrusive procedure for approximate likelihood construction and model discrepancy estimation. Lastly, the proposed BNOR framework is validated on the wave propagation problem in heterogeneous bars. It is found that the learnt model has 1) captured the correct band gap and the group velocity; 2) reproduced high-fidelity data for a composite material in applications that are substantially different from the training data under a significant range of conditions; 3) provided a characterization for the posterior distribution of the parameters as well as the confidence region for further prediction tasks.

In this work, we have focused on the high-fidelity measurements so the modeling discrepancy acts as the main contributor of uncertainties. As a natural follow-up, we plan to consider noisy observational data, for which a more sophisticated error model would be required to provide a principled way of attributing predictive uncertainties to components due to data noise/error, modeling discrepancy, as well as any additional errors associated with the lack-of-information. Another interesting future direction would be to incorporate our model while enabling targeted model improvement and optimal experimental design. Last but not least, to illustrate the efficiency of our algorithm, two- and three-dimensional test cases will also be considered. For example, one may use the BNOR framework to characterize the homogenization error in the coarse-grained nonlocal model from molecular dynamics simulations, as an extension of the development in \cite{you2021md}.

\subsection{Acknowledgments}

Y. Fan and Y. Yu would like to acknowledge support by the National Science Foundation under award DMS-1753031 and
the AFOSR grant FA9550-22-1-0197. Portions of this research were conducted on Lehigh University's Research Computing infrastructure partially supported by NSF Award 2019035. 
 
S. Silling and M. D'Elia would like to acknowledge the support of the Sandia National Laboratories (SNL) Laboratory-directed Research and Development program and by the U.S. Department of Energy (DOE), Office of Advanced Scientific Computing Research (ASCR) under the Collaboratory on Mathematics and Physics-Informed Learning Machines for Multiscale and Multiphysics Problems (PhILMs) project. H. Najm acknowledges support by the U.S. DOE ASCR office, Scientific Discovery through Advanced Computing (SciDAC) program. 
This article has been authored by an employee of National Technology and Engineering Solutions of Sandia, LLC under Contract No. DE-NA0003525 with the U.S. Department of Energy (DOE). The employee owns all right, title and interest in and to the article and is solely responsible for its contents. The United States Government retains and the publisher, by accepting the article for publication, acknowledges that the United States Government retains a non-exclusive, paid-up, irrevocable, world-wide license to publish or reproduce the published form of this article or allow others to do so, for United States Government purposes. The DOE will provide public access to these results of federally sponsored research in accordance with the DOE Public Access Plan https://www.energy.gov/downloads/doe-public-access-plan.

%

\end{document}